\documentclass[twocolumn,
preprintnumbers,prl,
superscriptaddress
]{revtex4-1}

\usepackage{amsmath,amssymb}
\usepackage[usenames,dvipsnames,svgnames,table]{xcolor}
\usepackage[colorlinks,citecolor=FireBrick,linkcolor=FireBrick,urlcolor=FireBrick,linktocpage]{hyperref}
\usepackage{graphicx}
\usepackage{times}
\usepackage{braket}
\usepackage{mathtools}
\usepackage{tikz}
\usetikzlibrary {decorations.pathmorphing}

\usepackage{mathrsfs}
\usetikzlibrary{arrows}
\usepackage{dsfont}
\usepackage{amsfonts}
\usepackage{amsmath,amssymb}
\usepackage{physics}
\usepackage{amsthm}
\usepackage{setspace}
\usepackage{relsize}
\usepackage{todonotes}
\usepackage{xpatch}

\newcommand{\CW}{\mathcal{W}}
\newcommand{\CL}{\mathcal{L}}
\newcommand{\CH}{\mathcal{H}}
\newcommand{\CD}{\mathcal{D}}
\newcommand{\CE}{\mathcal{E}}

\newcommand{\CC}{\mathcal{C}}

\newcommand{\CT}{\mathcal{T}}
\newcommand{\CI}{\mathcal{I}}

\newcommand{\CS}{\mathcal{S}}

\newcommand{\CM}{\mathcal{M}}
\newcommand{\CP}{\mathcal{P}}
\newcommand{\CV}{\mathcal{V}}

\newcommand{\DZ}{\mathds{Z}}

\newcommand{\be}{\begin{equation}}
\newcommand{\ee}{\end{equation}}
\newcommand{\bea}{\begin{eqnarray}}
\newcommand{\eea}{\end{eqnarray}}

\begin{document}

\title{Unveiling dynamical quantum error correcting codes via non-invertible symmetries}

\author{Rajath Radhakrishnan}
\affiliation{International Centre for Theoretical Physics, Strada Costiera 11, Trieste 34151, Italy}
\affiliation{Mathematical Institute, University of Oxford,
Andrew Wiles Building, Woodstock Road, Oxford, OX2 6GG, UK}
\author{Adar Sharon}
\affiliation{Simons Center for Geometry and Physics,
Stony Brook University, Stony Brook, NY, 11790, United States}
\affiliation{Mani L. Bhaumik Institute for Theoretical Physics, Department of Physics and Astronomy,
University of California, Los Angeles, CA 90095, USA}
\author{Nathanan Tantivasadakarn}
\affiliation{C. N. Yang Institute for Theoretical Physics, Stony Brook University, Stony Brook, NY 11794, USA}


\begin{abstract}
Dynamical stabilizer codes (DSCs) have recently emerged as a powerful generalization of static stabilizer codes for quantum error correction, replacing a fixed stabilizer group with a sequence of non-commuting measurements. This dynamical structure unlocks new possibilities for fault tolerance but also introduces new challenges, as errors must now be tracked across both space and time. In this work, we provide a physical and topological understanding of DSCs by establishing a correspondence between qudit Pauli measurements and non-invertible symmetries in 4+1-dimensional 2-form gauge theories. Sequences of measurements in a DSC are mapped to a fusion of the operators implementing these non-invertible symmetries. We show that the error detectors of a DSC correspond to endable surface operators in the gauge theory, whose endpoints define line operators, and that detectable errors are precisely those surface operators that braid non-trivially with these lines. Finally, we demonstrate how this framework naturally recovers the spacetime stabilizer code associated with a DSC. 
\end{abstract}

\maketitle

\section{Introduction}

Quantum error correction is essential for building a fault-tolerant quantum computer. Among the most studied constructions are quantum stabilizer codes \cite{Calderbank97,Gottesman:1997zz,Gottesman:1998se}. Recent work has emphasized the role of dynamical stabilizer codes (DSCs), obtained by evolving a stabilizer code through a sequence of measurements \cite{hastings2021dynamically,Haah2022boundarieshoneycomb,Aasen22,Gidney2021faulttolerant,Davydova23,Kesselring:2022eax,Zhang23,Bombin2024,Gidney2023,Bauer2024topologicalerror,Ellison23,Davydova:2023mnz,Dua24,Townsend-Teague23,Kobayashi:2023zxs,Higgott24,GransSamuelsson2024improvedpairwise,Fahimniya25,Bauer2025lowoverheadnon,Bauer2025x+y,MdlF25,Xu:2025zos}. In a DSC, the code subspace depends on the measurement sequence, and these codes exhibit several desirable features: stabilizers with low-weight Pauli operators and the ability to natively and fault-tolerantly implement logical operations. However, their dynamical nature makes error detection and correction more challenging. Unlike static stabilizer codes, one must adopt a spacetime perspective of errors, keeping track of both their spatial location and temporal occurrence \cite{Delfosse:2023aqk,Fu:2024hin,Blackwell25}. 

Quantum error correcting codes are closely related to quantum phases of matter and associated topological quantum field theories (TQFTs). The code subspace of a stabilizer code can be regarded as the ground-state Hilbert space of a quantum many-body system whose Hamiltonian is given by the stabilizer elements \cite{Kitaev:1997wr,Bombin:2013osf}. Moreover, salient features of the code, such as the number of logical qubits and types of errors, are already encoded in the continuum TQFTs describing the low-energy limit of the quantum many-body system. More recently, it was realized that one can construct asymptotically good Quantum LDPC codes \cite{tillich2013quantum,bravyi2014homological,hastings2021fiber,Panteleev:2021wvc,Breuckmann:2020jpn}. However, these codes cannot be embedded into Euclidean lattices in low dimensions \cite{bravyi2009no,bravyi2010tradeoffs}. Therefore, their interpretation as local quantum many-body systems is unclear, though interesting advancements have been made in this direction \cite{Yin:2024hbx,DeRoeck:2024alc}. This motivates a broader question: to what extent can general stabilizer codes be understood directly in terms of continuum field theories? Furthermore, can measurements and dynamical stabilizer codes (DSCs) also be naturally captured within this field-theoretic framework? 

In this paper, we given an answer to this question by establishing a one-to-one correspondence between qudit stabilizer codes and non-anomalous 2-form symmetries of a class of 4+1d TQFTs \footnote{The primary operators of certain 1+1d RCFTs and the corresponding line operators of their associated 3d Chern-Simons theories capture the properties of some static qudit stabilizer codes. However, this setting is limited to special stabilizer codes. This boils down to the fact that the braiding of line operators in 2+1d is symmetric, while the abelianized Pauli group has a symplectic structure \protect\cite{Dymarsky:2020qom,Dymarsky:2021xfc,Buican:2021uyp,Buican:2023ehi}.}. These symmetries are implemented by topological 2-dimensional surface operators in the TQFT. Under this correspondence, the measurement of a Pauli operator is realized as a 4-dimensional topological operator. The effect of a measurement on the stabilizer and logical operators of the code can be obtained from the action of the corresponding 4-dimensional operator on topological surface operators. The irreversibility of a measurement is directly related to the fact that these 4-dimensional operators implement non-invertible symmetries of the TQFT. 

A sequence of measurements in a dynamical stabilizer code corresponds to the sequential action of non-invertible symmetries on 2-dimensional surface operators. This viewpoint provides a new topological framework for understanding error correction in DSCs. In particular, we show that the detectors of a DSC are precisely those surface operators that can topologically end on two 4-dimensional operators associated with the measurements. Because of their topological nature, these endable surface operators can be continuously shrunk to line operators. In contrast, the logical and error operators of a DSC correspond to non-endable surface operators that cannot terminate on measurement operators. Errors detectable by the code are characterized by surface operators that braid non-trivially with the line operators obtained from endable surfaces. Finally, we relate this construction to the spacetime stabilizer code associated with a DSC \cite{Delfosse:2023aqk}: endable surface operators must mutually commute and correspond exactly to the stabilizers of the spacetime code.

\section{Error Correction with Dynamical Stabilizer Codes}

In this section, we will briefly review stabilizer codes and fix notation. We will define dynamical stabilizer codes and their detectors. We refer the readers to \cite{Calderbank97,Gottesman:1998se,farinholt2014ideal,haah2016algebraic} for more details.

\subsection{Stabilizer codes}

Consider a qudit with $p$-dimensional Hilbert space $\CH$, for some positive integer $p$. Let $X,Z$ be the generalized Pauli operators acting on this Hilbert space as
\be
X\ket{i}=\ket{i+1}~,~ Z\ket{j}=e^{\frac{2\pi ij}{p}}\ket{j}~.
\ee
The qudit Pauli group $\CP$ has elements $\omega X^aZ^b$, where $\omega$ is a $d^{\rm th}$ root of unity when $p$ is odd or $2p^{\rm th}$ root of unity otherwise. They satisfy the commutation relation
\be
X^aZ^b X^{a'}Z^{b'}=e^{\frac{2\pi i (ab'-ba')}{p}} X^{a'}Z^{b'}X^aZ^b ~.
\ee
For $n$ qudits, the generalized Pauli group $\CP_n$ acts on the $p^n$-dimensional Hilbert space $\CH_n$ and has elements of the form 
\be
G(\vec a,\vec b):= \omega^{\lambda} X^{a_1}Z^{b_1} \otimes \dots X^{a_n} Z^{b_n}~,
\ee
where $\vec a, \vec b \in \DZ_p^{2}$ and $\lambda \in \DZ_{p}$ for $p$ odd and $\lambda \in \DZ_{2p}$ for $p$ even. The commutation relation is
\be
G(\vec a,\vec b)G(\vec a',\vec b')= e^{\frac{2\pi i (\vec a,\vec b)*(a',\vec b')}{p}} G(\vec a',\vec b')G(\vec a,\vec b),
\ee
where the product $*$ is defined as 
\be
\label{eq:symplectic product}
(\vec a,\vec b)*(a',\vec b') = \sum_{i=1}^n a_ib_i'-a_i'b_i \text{ mod } p~. 
\ee
On quotienting with the subgroup $\langle \omega \rangle$ of $\CP_n$ generated by $\omega$, we get the abelianized Pauli group $\CV_n:=\CP_n/\langle \omega \rangle$. As a group $\CV_n$ is isomorphic to $\DZ_p^n \times \DZ_p^n$. The product $*$ in \eqref{eq:symplectic product} defines a symplectic inner product structure on $\CV_n$. A Stabilizer code is defined by an abelian subgroup $\CS$ of $\CP_n$. Equivalently, $\CS$ is an isotropic subgroup of $\CV_n$. The code subspace $\CC \subset \CH_n$ is the simultaneous $+1$ eigenspace of the generalised Pauli operators in $\CS$. The non-trivial logical operators acting on the encoded qubits are $\CL:= Z(\CS)/\CS$, where $Z(S)$ is the subgroup of elements in $\CP_n$ which commute with $\CS$. A Pauli operator $E$ that does not commute with at least one stabilizer element is a detectable error. Let $\CE$ be the set of equivalence classes of errors that can be detected using this code. For a given $d$ and $n$, let $\Pi_{p,n}$ be the set of all stabilizer groups. A choice of stabilizer group in this set defines a code with parameters $[[n,k,d]]$ where $k:=\text{dim}(\CC)$ is the number of encoded qubits and $d$ is equal to the minimum weight of a logical operator. $d$ is a measure of how good the code is in correcting errors. 

The quintessential example of a stabilizer code is the Toric code \cite{Kitaev:1997wr}. In its simplest instance, it can be regarded as a [[4,2,2]] code with stabilizer groups
\be
\CS=\langle XXXX, ZZZZ \rangle~.
\ee
This is the smallest code that can detect all single-qubit errors as any Pauli operator in $\CP_4$ acting on a single qubit anti-commutes with at least one of the stabilizer generators. Consider a sequence of time steps starting from $t=0$ when we measure the Pauli operators $XXXX$ and $ZZZZ$. Now, the state of the system is in the code subspace for the stabilizer group $\langle \pm 1 XXXX, \pm 1 ZZZZ\rangle$, where the signs are fixed by the measurement outcome. At time $t=1$, we measure the stabilizers again. If no error has occurred, then the outcome of this measurement is deterministically equal to the measurement outcomes at time $t=0$. However, if a single qubit error has occurred between $t=0$ and $t=1$, we can detect it. In a stabilizer code, the stabilizers need to be measured at frequent intervals to detect potential errors. 

\subsection{Dynamical stabilizer codes}

While the [[4,2,2]] above can detect single-qubit errors, it involves measuring 4-qubit operators. In several practical architectures, 2-qubit measurements are the most natural and easiest to implement~\cite{Aasen16,Bartolucci2023fusion}. Therefore, it is desirable to reduce the multi-qubit measurements into a sequence of at most 2-qubit measurements. To achieve this, we can use a sequence of non-commuting measurements. As an example, consider the 3-qubit measurement $ZZZ$.  It can be indirectly measured by making the following sequence of 2-qubit and 1-qubit measurements. 
\bea
&&\CM_1=\langle Z_1Z_2,Z_3Z_4,Z_5Z_6\rangle~,~\\
&&\CM_2=\langle X_2X_4,X_4X_5\rangle~,~\CM_3=\langle Z_2,Z_4,Z_6\rangle~.
\eea
On measuring the operators in round 1, we get a state which is a simultaneous eigenstate of $O(Z_1Z_2)Z_1Z_2, ~O(Z_3Z_4)Z_3Z_4$ and $O(Z_5Z_6)Z_5Z_6$ where $O(M)$ is the $\pm 1$ measurement outcome on measuring the Pauli operator $M$. Therefore, the resulting state belongs to the code subspace defined by the stabilizer group $\CS_1:=\langle O(Z_1Z_2)Z_1Z_2, ~O(Z_3Z_4)Z_3Z_4,~O(Z_5Z_6)Z_5Z_6\rangle$. $\CS_1$ is called the Instantaneous Stabilizer Group (ISG) at time $t=1$. After the subsequent measurements, we get the ISGs
\bea
&\CS_1=\langle Z_1Z_2,Z_3Z_4,Z_5Z_6\rangle \\
& \to \CS_2=\langle X_2X_4,X_4X_5,Z_1Z_2Z_3Z_4Z_5Z_6\rangle\\
& \to S_3 =\langle Z_2,Z_4,Z_6,Z_1Z_3Z_5\rangle ~.
\eea
In the expression above, we have simplified the notation by omitting the measurement outcomes. This shows that the 3-qubit operator $Z_1Z_3Z_5$ can be measured using a sequence of three single-qubit and 2-qubit measurements on 6 qubits. In general, the measurement of any high-weight Pauli operators can be reduced to one and two-qubit measurements at the expense of adding auxiliary qubits and increasing the number of measurements~\cite{delaFuente:2024ttz}. This observation shows that the high-weight measurements involved in static stabilizer codes can be replaced with a sequence of non-commuting 1 and 2-qubit measurements. Since the measurements are non-commuting, the stabilizer group changes with each measurement. Therefore, replacing high-weight Pauli operators with a sequence of non-commuting low-weight Pauli measurements converts static stabilizer codes into dynamical codes. The stabilizers of the [[4,2,2]] code can also be converted to a sequence of 1 and 2-qubit measurements~\cite{Townsend-Teague23}, and various decompositions have been studied for the toric code~\cite{Chao20,Gidney2023,GransSamuelsson2024improvedpairwise}. 

Consider an $n$ qubit system at time $t=0$ initialized in a maximally mixed state whose stabilizer group $\CS_0$ is trivial. A Dynamical Stabilizer Code (DSC) is defined by a sequence of measurements $\CM_1, \CM_2, \dots$ at time $t=1,2,\dots$ \cite{hastings2021dynamically,Townsend-Teague23,Fu:2024hin}. On measuring some Pauli operators $M_t\in \CM_t$, the stabilizer group $\CS_{t-1}$ evolves as follows:
\begin{enumerate}
\item $M_{t} \in \CS_{t-1}$. In this case, the measurement of $M_t$ does not change $\CS_{t-1}$. 

\item $M_t \notin \CS_{t-1}$ and $M_t \in Z(\CS_{t-1})$. In this case, the measurement updates the stabilizer group as $\CS_{t-1} \to \CS_{t-1} \cup \{O(M_t) M_t\}$, where $O(M_t)\pm 1$ is the outcome of the measurement of $M_t$.

\item $M_t \notin \CS_{t-1}$ and $M_t \notin Z(\CS_{t-1})$. In this case, $M_t$ anti-commutes with some elements of $\CS_{t-1}$. Let $\{S_1,\dots ,S_k\}$ be a generating set of elements in $\CS_{t-1}$ which anti-commute with $M_t$. Then, we update the stabilizer group as $\{S_1,\dots ,S_k\} \to \{O(M_t) M_t, S_1 S_2, \dots ,S_1 S_k\}$ and leaving the other elements of $\CS$ unchanged.
\end{enumerate}
Let $\CC_t$ be the $+1$ eigenspace of the ISG  $\CS_t$. Note that under each measurement, the number of generators of the stabilizer group either stays constant or increases. This shows that at each time step, the dimension of the code subspace $\CC_t$ either remains the same or decreases. Let $t=T$ be the time after which dim$(\CC_{T+t})=\text{dim}(\CC_T)$ for all $t>T$. dim$(\CC_T)=2^k$ defines the number of logical qubits of the dynamical code. If the measurements $\CM_1,\CM_2,\dots$ are chosen arbitrarily, after time $T$ the code subspace will more than likely become $1$-dimensional. Therefore, we cannot encode any logical qubits. However, as long as the measurement pairs are chosen carefully, it can be ensured that the stabilizer group does not increase in size after a finite time $T$ \cite{Aasen2023pjk}. 

\noindent \textbf{Example:} Let us consider a simple DSC on four physical qubits with measurement sequence $\CM_1=\langle Z_1Z_2, Z_3Z_4\rangle, \CM_{2}=\langle X_1X_3,X_2X_4\rangle$ repeated indefinitely. The subscript on a Pauli operator labels the qubit on which the operator acts. In this case, the ISGs are given by 
\bea
&\CS_1=\langle Z_1Z_2, Z_3Z_4\rangle \to \CS_2=\langle
X_1X_3,X_2X_4, Z_1Z_2Z_3Z_4 \rangle \nonumber \\
&\to \CS_3=\langle Z_1Z_2,X_1X_2X_3X_4,Z_1Z_2Z_3Z_4\rangle \nonumber \\
& \to \CS_4 =\langle X_1X_3,X_1X_2X_3X_4,Z_1Z_2Z_3Z_4\rangle ~\dots
\eea
Therefore, we find 
\be
\label{eq:Bacon-Shor ISG}
\CS_t = \begin{cases} \langle Z_1Z_2, Z_3Z_4 \rangle ~ & t=1,   \\ \langle X_1X_3,X_1X_2X_3X_4,Z_1Z_2Z_3Z_4\rangle & t \text{ even},  \\
\langle Z_1Z_2,X_1X_2X_3X_4,Z_1Z_2Z_3Z_4\rangle   & t>1, t \text{ odd}.
\end{cases}
\ee
The size of the stabilizer group in this case increases over the first few measurements and becomes a constant. This code has a single logical qubit. This is the DSC version of the Bacon-Shor code discussed in \cite{Townsend-Teague23}.

\subsection{Detectors of DSC}

In static stabilizer codes, detectable errors are Pauli operators that anti-commutes with the stabilizers. This is because in the absence of errors  the outcome of measuring the stabilizers is determined. Therefore, any deviation from the expected outcome indicates that an error has occurred. In static stabilizer codes, a pair of consecutive stabilizers are called \textit{detectors} as a change in value between two time steps indicates an error has occurred. In a DSC, the stabilizer group evolves in time. Therefore, whether the DSC can detect an error depends on the commutation relation of the error with the ISG, which depends on the temporal location of the error. Therefore, identifying detectable errors in DSC must use a spacetime approach. 

Detectors of a DSC arise from redundant measurements as follows. Consider the measurement of a Pauli operator $M_t \in \CM_t$ at time $t$. If $M_t \in \CS_{t-1}$, the measurement outcome of $M_t$ is determined by the measurements in the previous rounds. In particular, let $S_1,\dots, S_{n-k}$ be the set of generators of $\CS_{t-1}$. Also, let $M_t= \prod_{i \in \sigma }  S_i$, where $\sigma$ is a subset of $\{1,\dots n-k\}$, be the decomposition of $M_t$ in terms of the stabilizer generators. Then, we have 
\be
O(M_t)= \prod_{i\in \sigma} O(S_i)~.
\ee
By going through the measurements $\CM_t$ over all time steps, one can determine all such relations. The operator $D:=M_t\prod_{i\in \sigma} S_i$ is a detector as its measurement outcome is determined in the absence of errors. Let $\CD$ be the set of detectors obtained in this way. Clearly, the product of two detectors is also a detector. Therefore, $\CD$ is a group. 

Let us determine the detectors in the Bacon-Shor DSC discussed in the previous section. From \eqref{eq:Bacon-Shor ISG} it is clear that the operator $Z_1Z_2Z_3Z_4$ is measured twice at consecutive even time steps while the operator $X_1X_2X_3X_4$ is measured twice at consecutive odd time steps. Therefore, we have
\bea
\label{eq:Bacon-Shor detectors}
&&O(Z_{1,t}Z_{2,t}) O(Z_{3,t}Z_{4,t}) \nonumber \\ && \hspace{1.7cm} =  O(Z_{1,t-2}Z_{2,t-2}) O(Z_{3,t-2}Z_{4,t-2})~,~t \text{ odd,} \nonumber \\
&&O(X_{1,t}Z_{3,t}) O(X_{2,t}X_{4,t}) \nonumber \\ && \hspace{1.4cm} = O(X_{1,t-2}X_{3,t-2}) O(X_{2,t-2}X_{4,t-2}) ~,~t \text{ even}. \nonumber \\
\eea
Note that in the above expression, we have included subscripts to indicate the times at which the Pauli operators are measured. An error occurring before a measurement flips the corresponding measurement outcome whenever it fails to commute with the measurement operator. Consequently, a sequence of errors occurring at different times is detectable if it violates the conditions in \eqref{eq:Bacon-Shor detectors}. As in static stabilizer codes, one must then map each detector violation to the corresponding equivalence class of errors. In the next section, we address this problem using the framework of 2-form gauge theory.

\section{2-form gauge theory}

The abelian 2-form gauge theory is a 4+1d Topological Quantum Field Theory (TQFT) $\CT$ defined by the action \cite{Banks:2010zn,Kapustin:2014gua,Chen:2021xuc}
\be
\label{eq:2-form gauge theory action}
\sum_{i,j=1}^{2r} \frac{K_{ij}}{4\pi} \int b^{i}db^{j}~,
\ee
where $K$ is an anti-symmetric non-degenerate matrix and $b^{i}$ are $U(1)$ gauge fields. This theory has surface operators $U_{a}=e^{\oint \sum_i a_i b^i}$ labelled by $a \in A:=\DZ^{2r}/K\DZ^{2r}$ and the braiding of surface operators is given by
\be
\beta(U_a,U_b)= e^{2 \pi i a^{T} K^{-1} b}~.
\ee
This is very similar to abelian Chern-Simons (CS) theory. However, there are many crucial differences. In abelian CS theory the braiding of line operators is symmetric. In particular, if the self-braiding of line operators is trivial, then the mutual braidings are also trivial. In contrast, in the 4+1d abelian 2-form gauge theory, the surface operators $U_{i}$ have trivial self-braiding, but non-trivial mutual braidings because $K$ is anti-symmetric. Therefore, $A$ is the 2-form symmetry group of the theory and $\beta$ is its anomaly. Unlike in 2+1d, the generators of this symmetry do not have any self-anomaly even though their mixed anomaly is non-trivial. Note that $\beta$ is non-degenerate due to the invertibility of the matrix $K$. In other words, $\beta(U_i,U_j)$ is a non-degenerate anti-symmetric bilinear form on the group $A$.

Let $B\subset A$ be a non-anomalous subgroup of $A$. In other words, $\beta|_B$ is trivial. Since the anomaly is trivial, we can gauge the symmetry $B$ to get a new TQFT $\CT/B$. On gauging $B$ the surface operators in $\CT$ which braid trivially with $B$ remain as genuine surface operators in $\CT/B$. On the other hand, those which braid non-trivially with $B$ become non-genuine surface operators in $\CT/B$. 

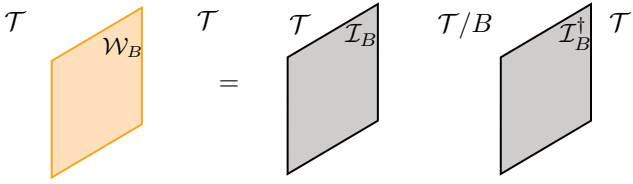
\begin{figure}[h!]
\centering

\tikzset{every picture/.style={line width=0.75pt}} 

\begin{tikzpicture}[x=0.75pt,y=0.75pt,yscale=-0.6,xscale=0.5]

\draw  [color={rgb, 255:red, 0; green, 0; blue, 0 }  ,draw opacity=1 ][fill={rgb, 255:red, 192; green, 188; blue, 188 }  ,fill opacity=0.75 ] (320.55,173) -- (320.57,74.86) -- (411.58,30.06) -- (411.57,128.19) -- cycle ;
\draw  [color={rgb, 255:red, 245; green, 166; blue, 35 }  ,draw opacity=1 ][fill={rgb, 255:red, 255; green, 224; blue, 187 }  ,fill opacity=1 ] (82.55,176) -- (82.57,77.86) -- (173.58,33.06) -- (173.57,131.19) -- cycle ;
\draw  [color={rgb, 255:red, 0; green, 0; blue, 0 }  ,draw opacity=1 ][fill={rgb, 255:red, 192; green, 188; blue, 188 }  ,fill opacity=0.75 ] (535.55,173) -- (535.57,74.86) -- (626.58,30.06) -- (626.57,128.19) -- cycle ;

\draw (130.86,55.93) node [anchor=north west][inner sep=0.75pt]    {\small $\CW_B$};
\draw (32,34.4) node [anchor=north west][inner sep=0.75pt]    {$\CT$};
\draw (226,33.4) node [anchor=north west][inner sep=0.75pt]    {$\CT$};
\draw (248,91.4) node [anchor=north west][inner sep=0.75pt]    {$=$};
\draw (319,36.4) node [anchor=north west][inner sep=0.75pt]    {$\CT$};
\draw (642,35.4) node [anchor=north west][inner sep=0.75pt]    {$\CT$};
\draw (376,44.4) node [anchor=north west][inner sep=0.75pt]    {$\CI_B$};
\draw (592,38.4) node [anchor=north west][inner sep=0.75pt]    {$\CI_B^{\dagger }$};
\draw (468.58,34.46) node [anchor=north west][inner sep=0.75pt]    {$\CT/B$};
\end{tikzpicture}
\caption{Fusing the interface $\CI_B$ with its dual $\CI_B^{\dagger}$ gives a 4-dimensional operator $\CW_B$ in the TQFT $\CT$.}
\label{fig:codimension 1 operator from I}
\end{figure}

Gauging $B$ can be used to construct 4-dimensional operators which will play an important role in our analysis of dynamical codes. To understand this construction, let us consider the interface $\CI_B$ between the TQFTs $\CT$ and $\CT/B$ which implements this gauging. As a map $\CI_B: \CT\to \CT/B$, it is given by
\be
\label{eq:action of I}
 \CI_B(U)=\begin{cases}\sum_{U'\in B} U' \times U~ ~ & \text{ if } \beta(U,U')=1 ~\forall~ U'\in B ,   \\0 &  \text{ otherwise} .\end{cases}
\ee
Here, $0$ denotes that passing $U$ through the interface $\CI$, results in a non-genuine surface operator in $\CT/B$. Using $\CI$, we can construct a 4-dimensional operator $\CW_B$ in $\CT$ given by \cite{Buican:2023bzl,KNBalasubramanian:2025vum} (see Fig. \ref{fig:codimension 1 operator from I})
\be
\label{eq:condensation defect}
\CW_B:=\CI_B^{\dagger} \CI_B: \CT \to \CT. 
\ee
This is the condensation defect obtained from higher-gauging the symmetry $B$ on a 4-dimensional subspace of of the 5-dimensional spacetime \cite{Roumpedakis:2022aik,Buican:2023bzl}. 

\section{DSC from Non-invertible Symmetries}

In this section, we will show that there exists an isomorphism between abelianized qudit Pauli group and the symmetry group $A$ of certain 2-form gauge theories. We will use this isomorphism to relate measurements of Pauli operators and non-invertible symmetries of the gauge theory. To that end, recall that the group $A$ of surface operators admit a non-degenerate anti-symmetric bilinear form $\beta$ through their braiding. Any such abelian group admits a Lagrangian decomposition \cite[Lemma 5.2]{davydov2007twisted} (see also \cite{Johnson-Freyd:2021tbq})
\be
A= D \oplus \hat D~,
\ee
where $\hat D$ is the group of irreducible representations of $D$.
Through this decomposition, any element of $A$ can be written in the form $(d,\pi)$ where $d\in D$ and $\pi \in \hat D$. The surface operators labelled purely by $D$ and $\hat D$ can be thought of as the fluxes and charges of the gauge group, respectively. In this labelling of the surface operators, the anomaly $\beta$ is given by 
\be
\beta(S_{(d,\pi)},S_{(d',\pi')})= \pi(d')\pi'(d)^{-1}~.
\ee
Suppose we choose $K$ such that the group $A$ has the decomposition 
\be
A= \DZ_p \times \hat \DZ_p~.
\ee
In other words, $A$ is a product of cyclic groups which is also its Lagrangian decomposition. Elements of $\hat \DZ_p$ are of the form $\pi_c(d)=e^{\frac{2\pi i c d}{p}}$ where $c,d \in \{0,\dots,p-1\}$. Therefore, the anomaly of the symmetry $A$ has the form
\be
\label{eq:anomaly simplified expression}
\beta(S_{(d,\pi_c)},S_{(d',\pi_{c'})})= e^{\frac{2 \pi i(cd'-c'd)}{p}}~.
\ee
Clearly, there is an isomorphism between the abelianized  generalized Pauli group $\CV$ acting on one qudit with $p$ states and the symmetry $A$ where the symplectic form \eqref{eq:symplectic product} is given by the anomaly $\beta$ \eqref{eq:anomaly simplified expression}. More generally, when 
\be
A= \DZ_p^n \times \hat \DZ_p^n~,
\ee
we have an isomorphism 
\be
\label{eq:main isomorphism}
\phi:  \CV_n \to A
\ee
between the abelianized generalized Pauli group $\CV_n$ acting on $n$ qudits and the symmetry group $A$ under which the symplectic form on $\CV_n$ is mapped to the anomaly $\beta$ of $A$. In fact, in general, the group $A$ can be written as $D \times \hat D$, where $D$ is a product of cyclic groups of different order. In this case, $A$ is an abelianized Pauli group acting on a system of qudits of varying dimensions. In the following, we will focus on $p=2$, even though the discussion applies more generally. 

\subsection{Measurements and non-invertible symmetries}

A subset of Pauli operators $\CM$ which commute with each other can be simultaneously measured. In fact, if commuting operators $M_1$ and $M_2$ are measured, the measurement outcome of $M_1M_2$ is also determined. Therefore, we can take a commuting subgroup $\CM$ of $\CP_n$ whose generators can be simultaneously measured. Under the isomorphism $\phi$, these are in one-to-one correspondence with non-anomalous subgroups of $A$. We can use this correspondence to study DSCs in terms of the 2-form gauge theory. To that end, let us first consider a static stabilizer code defined on $n$ qubits, which can be regarded as a DSC with measurement sequence $\CM_t=\CS, \forall t>0$ where $\CS$ is a abelian subgroup of the Pauli group. This defines a static stabilizer code with stabilizer group $\CS$. Under the isomorphism $\phi$, $\phi(\CS)$ is a non-anomalous subgroup of $A$. Conversely, each non-anomalous subgroup of $A$ corresponds to a static stabilizer code. Moreover, any 2-form gauge theory admits a gapped boundary obtained by gauging a maximal non-anomalous subgroup of A. When the 2-form gauge theory is interpreted as the SymTFT of a 4-dimensional QFT with a fixed symmetry \cite{Gaiotto:2020iye,Apruzzi:2021nmk,Chatterjee:2022kxb,Freed:2022qnc,Kaidi:2022cpf}, its different gapped boundaries correspond to the different possible ways of gauging the symmetry of the 4d theory. Through the isomorphism $\phi$, these gapped boundaries are in one-to-one correspondence with stabilizer codes that have a 1-dimensional code subspace.

\begin{figure}[h!]
\centering

\tikzset{every picture/.style={line width=0.75pt}} 

\begin{tikzpicture}[x=0.75pt,y=0.75pt,yscale=-0.7,xscale=0.7]

\draw  [color={rgb, 255:red, 245; green, 166; blue, 35 }  ,draw opacity=1 ][fill={rgb, 255:red, 255; green, 224; blue, 187 }  ,fill opacity=1 ] (269.82,171.73) -- (269.84,73.59) -- (360.86,28.79) -- (360.84,126.93) -- cycle ;
\draw [color={rgb, 255:red, 139; green, 6; blue, 24 }  ,draw opacity=1 ]   (316.99,101.91) -- (436,101.76) ;
\draw [color={rgb, 255:red, 139; green, 6; blue, 24 }  ,draw opacity=1 ] [dash pattern={on 4.5pt off 4.5pt}]  (270,102.76) -- (316.99,101.91) ;
\draw [color={rgb, 255:red, 139; green, 6; blue, 24 }  ,draw opacity=1 ]   (190.97,102.21) -- (265,102.76) ;
\draw  [color={rgb, 255:red, 139; green, 6; blue, 24 }  ,draw opacity=1 ][fill={rgb, 255:red, 139; green, 6; blue, 24 }  ,fill opacity=1 ] (313.69,101.91) .. controls (313.69,101) and (314.43,100.26) .. (315.34,100.26) .. controls (316.25,100.26) and (316.99,101) .. (316.99,101.91) .. controls (316.99,102.82) and (316.25,103.56) .. (315.34,103.56) .. controls (314.43,103.56) and (313.69,102.82) .. (313.69,101.91) -- cycle ;

\draw (308,55.4) node [anchor=north west][inner sep=0.75pt]    {$\CW_{\phi(\CS)}$};
\draw (439,92.4) node [anchor=north west][inner sep=0.75pt]    {$\CW_{\phi(\CS)}(U)$};
\draw (177,91.4) node [anchor=north west][inner sep=0.75pt]    {$U$};

\end{tikzpicture}
\caption{Action of the 4-dimensional operator $\CW_{\phi(\CS)}$ on 2-dimensional surface operator $U$. Note that we have drawn the diagram as a 2d operator acting on a 1d operator for clarity.}
\label{fig: action of W}	
\end{figure}
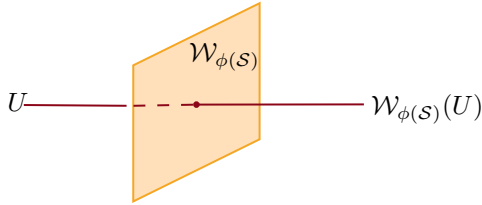

As a preparation to study general DSCs, it is useful to consider the 4-dimensional operator $\CW_{\phi(\CS)}$ defined in \eqref{eq:condensation defect}. Let us study the action of $\CW_{\phi(\CS)}$ on a general surface operator $U'$. Using \eqref{eq:action of I} and the definition of $\CW_{\phi(\CS)}$, we have (see Fig. \ref{fig: action of W})
\be
\label{eq:action of W}
 \CW_{\phi(\CS)}(U')=\begin{cases}\sum_{U\in \phi(\CS)} U \times U'~ ~ & \text{ if } \beta(U,U')=1 ,   \\0 &  \text{ otherwise} .\end{cases}
\ee
This shows that the surface operators which pass through the operator $\CW_{\phi(\CS)}$ correspond to the logical operators of the static stabilizer code defined by $\CS$. In fact, $\CW_{\phi(U)}(U')$ is a direct sum of operators which correspond to the equivalence class of Pauli operators which act on the code subspace as the same logical operator. In particular, we have
\be
 \CW_{\phi(\CS)}(\mathds{1})=\sum_{U\in \phi(\CS)} U~.
\ee
This shows that under the action of the  $\CW_{\phi(\CS)}$, the surface operators in $\phi(\CS)$ are identified with the trivial surface operator $\mathds{1}$. This corresponds to the fact that the Pauli operators in $\CS$ act trivially on the code subspace. Finally, there are surface operators which, on passing through the operator $\CW_{\phi(\CS)}$, become non-genuine surface operators. A surface operator $U' \in A$ become non-genuine on passing through $\CW_{\phi(\CS)}$ precisely when $\beta(U,U')\neq 1$ for some $U \in \phi(\CS)$. Therefore, these are the detectable errors of the static stabilizer code. 

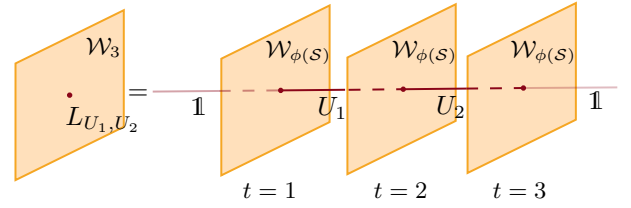
\begin{figure}[h!]
\centering

\tikzset{every picture/.style={line width=0.75pt}} 

\tikzset{every picture/.style={line width=0.75pt}} 

\begin{tikzpicture}[x=0.75pt,y=0.75pt,yscale=-0.6,xscale=0.6]

\draw  [color={rgb, 255:red, 245; green, 166; blue, 35 }  ,draw opacity=1 ][fill={rgb, 255:red, 255; green, 224; blue, 187 }  ,fill opacity=1 ] (250.13,217) -- (250.15,118.86) -- (341.17,74.06) -- (341.15,172.19) -- cycle ;
\draw  [color={rgb, 255:red, 245; green, 166; blue, 35 }  ,draw opacity=1 ][fill={rgb, 255:red, 255; green, 224; blue, 187 }  ,fill opacity=1 ] (77.13,221) -- (77.15,122.86) -- (168.17,78.06) -- (168.15,176.19) -- cycle ;
\draw  [color={rgb, 255:red, 139; green, 6; blue, 24 }  ,draw opacity=1 ][fill={rgb, 255:red, 139; green, 6; blue, 24 }  ,fill opacity=1 ] (121.23,149.53) .. controls (121.23,148.74) and (121.87,148.11) .. (122.65,148.11) .. controls (123.43,148.11) and (124.07,148.74) .. (124.07,149.53) .. controls (124.07,150.31) and (123.43,150.95) .. (122.65,150.95) .. controls (121.87,150.95) and (121.23,150.31) .. (121.23,149.53) -- cycle ;
\draw  [color={rgb, 255:red, 139; green, 6; blue, 24 }  ,draw opacity=1 ][fill={rgb, 255:red, 139; green, 6; blue, 24 }  ,fill opacity=1 ] (298.65,145.53) .. controls (298.65,144.74) and (299.28,144.11) .. (300.07,144.11) .. controls (300.85,144.11) and (301.48,144.74) .. (301.48,145.53) .. controls (301.48,146.31) and (300.85,146.95) .. (300.07,146.95) .. controls (299.28,146.95) and (298.65,146.31) .. (298.65,145.53) -- cycle ;
\draw  [color={rgb, 255:red, 139; green, 6; blue, 24 }  ,draw opacity=1 ][fill={rgb, 255:red, 139; green, 6; blue, 24 }  ,fill opacity=1 ] (443.07,137.53) .. controls (443.07,136.74) and (443.7,136.11) .. (444.48,136.11) .. controls (445.26,136.11) and (445.9,136.74) .. (445.9,137.53) .. controls (445.9,138.31) and (445.26,138.95) .. (444.48,138.95) .. controls (443.7,138.95) and (443.07,138.31) .. (443.07,137.53) -- cycle ;
\draw [color={rgb, 255:red, 139; green, 6; blue, 24 }  ,draw opacity=1 ]   (298.65,145.53) -- (354.17,145) ;
\draw  [color={rgb, 255:red, 245; green, 166; blue, 35 }  ,draw opacity=1 ][fill={rgb, 255:red, 255; green, 224; blue, 187 }  ,fill opacity=1 ] (356.13,216) -- (356.15,117.86) -- (447.17,73.06) -- (447.15,171.19) -- cycle ;
\draw  [color={rgb, 255:red, 245; green, 166; blue, 35 }  ,draw opacity=1 ][fill={rgb, 255:red, 255; green, 224; blue, 187 }  ,fill opacity=1 ] (457.13,215) -- (457.15,116.86) -- (548.17,72.06) -- (548.15,170.19) -- cycle ;
\draw  [color={rgb, 255:red, 139; green, 6; blue, 24 }  ,draw opacity=1 ][fill={rgb, 255:red, 139; green, 6; blue, 24 }  ,fill opacity=1 ] (401.65,144.53) .. controls (401.65,143.74) and (402.28,143.11) .. (403.07,143.11) .. controls (403.85,143.11) and (404.48,143.74) .. (404.48,144.53) .. controls (404.48,145.31) and (403.85,145.95) .. (403.07,145.95) .. controls (402.28,145.95) and (401.65,145.31) .. (401.65,144.53) -- cycle ;
\draw  [color={rgb, 255:red, 139; green, 6; blue, 24 }  ,draw opacity=1 ][fill={rgb, 255:red, 139; green, 6; blue, 24 }  ,fill opacity=1 ] (502.65,143.53) .. controls (502.65,142.74) and (503.28,142.11) .. (504.07,142.11) .. controls (504.85,142.11) and (505.48,142.74) .. (505.48,143.53) .. controls (505.48,144.31) and (504.85,144.95) .. (504.07,144.95) .. controls (503.28,144.95) and (502.65,144.31) .. (502.65,143.53) -- cycle ;
\draw [color={rgb, 255:red, 139; green, 6; blue, 24 }  ,draw opacity=1 ]   (404.48,144.53) -- (456,144) ;
\draw [color={rgb, 255:red, 139; green, 6; blue, 24 }  ,draw opacity=1 ] [dash pattern={on 4.5pt off 4.5pt}]  (456,144) -- (503.48,143.53) ;
\draw [color={rgb, 255:red, 139; green, 6; blue, 24 }  ,draw opacity=1 ] [dash pattern={on 4.5pt off 4.5pt}]  (357.13,145.06) -- (401.65,144.53) ;
\draw [color={rgb, 255:red, 139; green, 6; blue, 24 }  ,draw opacity=0.27 ]   (505.48,143.53) -- (583,143) ;
\draw [color={rgb, 255:red, 139; green, 6; blue, 24 }  ,draw opacity=0.43 ] [dash pattern={on 4.5pt off 4.5pt}]  (248,146) -- (298.65,145.53) ;
\draw [color={rgb, 255:red, 139; green, 6; blue, 24 }  ,draw opacity=0.37 ]   (192.48,146.53) -- (248,146) ;

\draw (133.86,98.93) node [anchor=north west][inner sep=0.75pt]    {\small $\CW_{3}$};
\draw (169,142.4) node [anchor=north west][inner sep=0.75pt]    {$=$};
\draw (116.65,157.35) node [anchor=north west][inner sep=0.75pt]    {$L_{U_{1}, U_{2}}$};
\draw (328.41,148.66) node [anchor=north west][inner sep=0.75pt]    {$U_{1}$};
\draw (428.24,147.66) node [anchor=north west][inner sep=0.75pt]    {$U_{2}$};
\draw (284.86,100.93) node [anchor=north west][inner sep=0.75pt]    {\small $\CW_{\phi(\CS)}$};
\draw (390.86,100.93) node [anchor=north west][inner sep=0.75pt]    {\small $\CW_{\phi(\CS)}$};
\draw (490.86,100.93) node [anchor=north west][inner sep=0.75pt]    {\small $\CW_{\phi(\CS)}$};
\draw (222.24,149.66) node [anchor=north west][inner sep=0.75pt]    {$\mathds{1}$};
\draw (556.24,144.66) node [anchor=north west][inner sep=0.75pt]    {$\mathds{1}$};
\draw (265.86,220.93) node [anchor=north west][inner sep=0.75pt]    {\small $t=1$};
\draw (375.86,220.93) node [anchor=north west][inner sep=0.75pt]    {\small $t=2$};
\draw (475.86,220.93) node [anchor=north west][inner sep=0.75pt]    {\small $t=3$};
\end{tikzpicture}
\caption{Consider surface operators $U_1$ and $U_2$ which can from junction with the trivial surface operator $\mathds{1}$ as in this figure. On fusing the three operators $\CW_{\phi(\CS)} $, we get a line operator $L_{U_1,U_2}$ on $\CW_3$.}
\label{fig:lines from endable surfaces}	
\end{figure}

In fact, errors are detected not from a single measurement of the stabilizers, but rather measurements of stabilizer at frequent intervals. Let us consider the measurement of the stabilizer group $\CS$ at time $t=1,2,3$. In the absence of errors, the result of measuring the stabilizer group $\CS$ at time $t$ is completely determined by the measurement at time $t-1$. Therefore, a Pauli error that occurs between time $t-1$ and $t$ can be detected if it anticommutes with $\CS$. Now, consider three 4-dimensional operators $\CW_{\phi(\CS)}$ corresponding to measuring $\CS$ at $t=1,2,3$. Consider the configuration in Fig. \ref{fig:lines from endable surfaces}.  Using \eqref{eq:action of W}, we know that this configuration of topological operators is consistent if and only if $U_1,U_2 \in \phi(\CS)$. We say that the surface operators $U_1,U_2$ are \textit{endable} on the operator $\CW_{\phi(\CS)}$. On parallel fusion of the three operators, the configuration of surface operators becomes a line operator $L_{U_1,U_2}$ on the operator $\CW_3:=\CW_{\phi(\CS)} \times \CW_{\phi(\CS)} \times \CW_{\phi(\CS)}$. Since $U_1,U_2\in \phi(\CS)$, we know that
\be
\beta(U_1,U_2)=1~.
\ee
This is consistent with the fact that the resulting line operator $L_{U_1,U_1}$ must braid trivially with itself as topological line operators cannot have non-trivial linking on the 4-dimensional operator $\CW_3$. For different choices of $U_1,U_2 \in \phi(\CS)$, we get different line operators on $\CW_3$ which all braid trivially with each other. 

The line operators $L_{U_1,U_2}$ are the detectors of this code, and they detect errors through non-trivially braiding with the surface operators on the operator $\CW_3$. To see this, consider some surface operators $V_1,V_2$ as in Fig. \ref{fig:surfaces of W3}. On parallel fusion of the three $\CW_{\phi(\CS)}$ operators, we get a surface operator $J_{V_1,V_2}$ on the operator $\CW_3$. Every surface operator on $\CW_3$ arises in this way. Now, consider some detectable error $E$ that occurs just before time $t=3$ as in Fig. \ref{fig:detectable errors from linking}. This corresponds to $V_1=\mathds{1}, V_2:=\phi(E)$. On fusing the three $\CW_{\phi(\CS)}$ operators, we get a surface operator, say $J_{\mathds{1},\phi(E)}$ on $\CW_3$. Since $E$ is a detectable error, there is some stabilizer element $S \in \CS$ that does not commute with it. Choose $U_2=\phi(S)$. Then, unlinking the surface operators $U_2$ and $V_2$ will produce a non-trivial phase. Therefore, unlinking the surface operator $J_{\mathds{1},\phi(E)}$ and the line operator $L_{U_1,U_2}$ will also produce a non-trivial phase. Conversely, every surface operator on $\CW_3$ that braids non-trivially with some $L_{U_1,U_2}$ corresponds to a sequence of errors $V_1,V_2$ that can be detected by the stabilizer code. In general, we can consider a sequence of several $\CW_{\phi(\CS)}$ operators corresponding to the measurements at time $t=1$ to $\Delta$. Fusing the operators gives the operator $\CW_{\Delta}$. The line operators $L_{U_1,\dots, U_{\Delta-1}}$ on $\CW_\Delta$ defined similar to the configuration in Fig. \ref{fig:lines from endable surfaces} are the detectors which can detect a pattern of errors $V_1,\dots, V_{\Delta-1}$ if the surface operator $J_{V_1,\dots,V_{\Delta-1}}$ braids non-trivially with $L_{U_1,\dots,U_{\Delta-1}}$. 
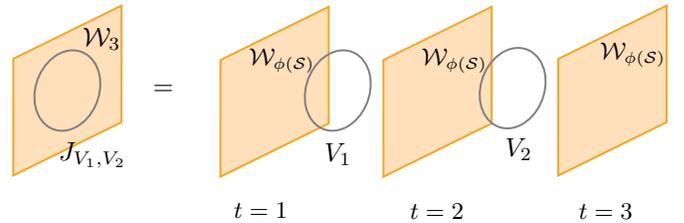
\begin{figure}[h!]
\centering

\tikzset{every picture/.style={line width=0.75pt}} 

\begin{tikzpicture}[x=0.75pt,y=0.75pt,yscale=-0.6,xscale=0.6]

\draw  [color={rgb, 255:red, 245; green, 166; blue, 35 }  ,draw opacity=1 ][fill={rgb, 255:red, 255; green, 224; blue, 187 }  ,fill opacity=1 ] (234.13,216) -- (234.15,117.86) -- (325.17,73.06) -- (325.15,171.19) -- cycle ;
\draw  [color={rgb, 255:red, 245; green, 166; blue, 35 }  ,draw opacity=1 ][fill={rgb, 255:red, 255; green, 224; blue, 187 }  ,fill opacity=1 ] (59.98,214.97) -- (60,116.83) -- (151.02,72.03) -- (151,170.17) -- cycle ;
\draw  [color={rgb, 255:red, 139; green, 6; blue, 24 }  ,draw opacity=1 ][fill={rgb, 255:red, 139; green, 6; blue, 24 }  ,fill opacity=1 ] (458.07,137.53) .. controls (458.07,136.74) and (458.7,136.11) .. (459.48,136.11) .. controls (460.26,136.11) and (460.9,136.74) .. (460.9,137.53) .. controls (460.9,138.31) and (460.26,138.95) .. (459.48,138.95) .. controls (458.7,138.95) and (458.07,138.31) .. (458.07,137.53) -- cycle ;
\draw  [color={rgb, 255:red, 245; green, 166; blue, 35 }  ,draw opacity=1 ][fill={rgb, 255:red, 255; green, 224; blue, 187 }  ,fill opacity=1 ] (371.13,216) -- (371.15,117.86) -- (462.17,73.06) -- (462.15,171.19) -- cycle ;
\draw  [color={rgb, 255:red, 245; green, 166; blue, 35 }  ,draw opacity=1 ][fill={rgb, 255:red, 255; green, 224; blue, 187 }  ,fill opacity=1 ] (518.13,215) -- (518.15,116.86) -- (609.17,72.06) -- (609.15,170.19) -- cycle ;
\draw  [color={rgb, 255:red, 128; green, 128; blue, 128 }  ,draw opacity=1 ] (305.92,156.91) .. controls (302.24,139.77) and (311.28,119.85) .. (326.13,112.43) .. controls (340.97,105.01) and (355.99,112.9) .. (359.67,130.05) .. controls (363.36,147.19) and (354.31,167.11) .. (339.47,174.53) .. controls (324.63,181.95) and (309.61,174.06) .. (305.92,156.91) -- cycle ;
\draw  [color={rgb, 255:red, 128; green, 128; blue, 128 }  ,draw opacity=1 ] (452.92,154.91) .. controls (449.24,137.77) and (458.28,117.85) .. (473.13,110.43) .. controls (487.97,103.01) and (502.99,110.9) .. (506.67,128.05) .. controls (510.36,145.19) and (501.31,165.11) .. (486.47,172.53) .. controls (471.63,179.95) and (456.61,172.06) .. (452.92,154.91) -- cycle ;
\draw  [color={rgb, 255:red, 128; green, 128; blue, 128 }  ,draw opacity=1 ] (78.63,156.93) .. controls (74.94,139.79) and (83.99,119.87) .. (98.83,112.45) .. controls (113.67,105.03) and (128.69,112.92) .. (132.37,130.07) .. controls (136.06,147.21) and (127.01,167.13) .. (112.17,174.55) .. controls (97.33,181.97) and (82.31,174.08) .. (78.63,156.93) -- cycle ;

\draw (115.86,87.93) node [anchor=north west][inner sep=0.75pt]    {$\CW_{3}$};
\draw (175,136.4) node [anchor=north west][inner sep=0.75pt]    {$=$};
\draw (95.65,184.35) node [anchor=north west][inner sep=0.75pt]    {$J_{V_{1}, V_{2}}$};
\draw (255.86,105.93) node [anchor=north west][inner sep=0.75pt]    {\small $\CW_{\phi(\CS)}$};
\draw (400.86,105.93) node [anchor=north west][inner sep=0.75pt]    {\small $\CW_{\phi(\CS)}$};
\draw (550.86,100.93) node [anchor=north west][inner sep=0.75pt]    {\small $\CW_{\phi(\CS)}$};
\draw (319,185.4) node [anchor=north west][inner sep=0.75pt]    {$V_{1}$};
\draw (471,181.4) node [anchor=north west][inner sep=0.75pt]    {$V_{2}$};
\draw (243,235.4) node [anchor=north west][inner sep=0.75pt]    {\small $t=1$};
\draw (533,236.4) node [anchor=north west][inner sep=0.75pt]    {\small $t=3$};
\draw (391,236.4) node [anchor=north west][inner sep=0.75pt]    {\small $t=2$};

\end{tikzpicture}
\caption{A surface operator on $\CW_3$ can be obtained from a non-unqiue configuration of surface operators $V_1$ and $V_2$ of the TQFT $\CT$.}
\label{fig:surfaces of W3}
\end{figure}
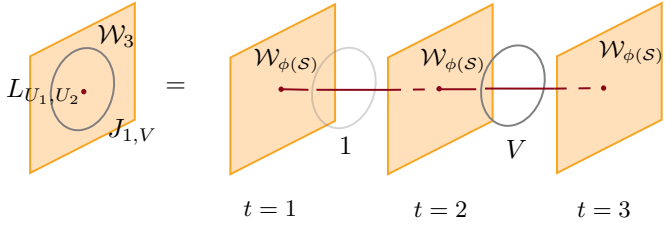
\begin{figure}[h!]
\centering

\tikzset{every picture/.style={line width=0.75pt}} 

\begin{tikzpicture}[x=0.75pt,y=0.75pt,yscale=-0.6,xscale=0.58]

\draw  [color={rgb, 255:red, 245; green, 166; blue, 35 }  ,draw opacity=1 ][fill={rgb, 255:red, 255; green, 224; blue, 187 }  ,fill opacity=1 ] (228.13,204) -- (228.15,105.86) -- (319.17,61.06) -- (319.15,159.19) -- cycle ;
\draw  [color={rgb, 255:red, 245; green, 166; blue, 35 }  ,draw opacity=1 ][fill={rgb, 255:red, 255; green, 224; blue, 187 }  ,fill opacity=1 ] (53.98,202.97) -- (54,104.83) -- (145.02,60.03) -- (145,158.17) -- cycle ;
\draw  [color={rgb, 255:red, 139; green, 6; blue, 24 }  ,draw opacity=1 ][fill={rgb, 255:red, 139; green, 6; blue, 24 }  ,fill opacity=1 ] (452.07,125.53) .. controls (452.07,124.74) and (452.7,124.11) .. (453.48,124.11) .. controls (454.26,124.11) and (454.9,124.74) .. (454.9,125.53) .. controls (454.9,126.31) and (454.26,126.95) .. (453.48,126.95) .. controls (452.7,126.95) and (452.07,126.31) .. (452.07,125.53) -- cycle ;
\draw  [color={rgb, 255:red, 245; green, 166; blue, 35 }  ,draw opacity=1 ][fill={rgb, 255:red, 255; green, 224; blue, 187 }  ,fill opacity=1 ] (365.13,204) -- (365.15,105.86) -- (456.17,61.06) -- (456.15,159.19) -- cycle ;
\draw  [color={rgb, 255:red, 245; green, 166; blue, 35 }  ,draw opacity=1 ][fill={rgb, 255:red, 255; green, 224; blue, 187 }  ,fill opacity=1 ] (512.13,203) -- (512.15,104.86) -- (603.17,60.06) -- (603.15,158.19) -- cycle ;
\draw  [color={rgb, 255:red, 128; green, 128; blue, 128 }  ,draw opacity=0.33 ] (299.92,144.91) .. controls (296.24,127.77) and (305.28,107.85) .. (320.13,100.43) .. controls (334.97,93.01) and (349.99,100.9) .. (353.67,118.05) .. controls (357.36,135.19) and (348.31,155.11) .. (333.47,162.53) .. controls (318.63,169.95) and (303.61,162.06) .. (299.92,144.91) -- cycle ;
\draw  [color={rgb, 255:red, 128; green, 128; blue, 128 }  ,draw opacity=1 ] (446.92,142.91) .. controls (443.24,125.77) and (452.28,105.85) .. (467.13,98.43) .. controls (481.97,91.01) and (496.99,98.9) .. (500.67,116.05) .. controls (504.36,133.19) and (495.31,153.11) .. (480.47,160.53) .. controls (465.63,167.95) and (450.61,160.06) .. (446.92,142.91) -- cycle ;
\draw  [color={rgb, 255:red, 128; green, 128; blue, 128 }  ,draw opacity=1 ] (72.81,145.78) .. controls (69.02,128.16) and (77.89,107.92) .. (92.62,100.56) .. controls (107.35,93.19) and (122.36,101.5) .. (126.15,119.11) .. controls (129.93,136.72) and (121.06,156.97) .. (106.33,164.33) .. controls (91.6,171.7) and (76.59,163.39) .. (72.81,145.78) -- cycle ;
\draw  [color={rgb, 255:red, 139; green, 6; blue, 24 }  ,draw opacity=1 ][fill={rgb, 255:red, 139; green, 6; blue, 24 }  ,fill opacity=1 ] (99.23,134.53) .. controls (99.23,133.74) and (99.87,133.11) .. (100.65,133.11) .. controls (101.43,133.11) and (102.07,133.74) .. (102.07,134.53) .. controls (102.07,135.31) and (101.43,135.95) .. (100.65,135.95) .. controls (99.87,135.95) and (99.23,135.31) .. (99.23,134.53) -- cycle ;
\draw [color={rgb, 255:red, 139; green, 6; blue, 24 }  ,draw opacity=1 ]   (303,133) -- (366.13,133.06) ;
\draw [color={rgb, 255:red, 139; green, 6; blue, 24 }  ,draw opacity=1 ]   (449.48,131.95) -- (510.17,132) ;
\draw [color={rgb, 255:red, 139; green, 6; blue, 24 }  ,draw opacity=1 ] [dash pattern={on 4.5pt off 4.5pt}]  (510.17,132) -- (554.65,131.53) ;
\draw [color={rgb, 255:red, 139; green, 6; blue, 24 }  ,draw opacity=1 ] [dash pattern={on 4.5pt off 4.5pt}]  (366.13,133.06) -- (410.65,132.53) ;
\draw [color={rgb, 255:red, 139; green, 6; blue, 24 }  ,draw opacity=1 ]   (410.65,132.53) -- (441.91,132.53) ;
\draw [color={rgb, 255:red, 139; green, 6; blue, 24 }  ,draw opacity=1 ]   (273.65,132.53) -- (296,133) ;
\draw  [color={rgb, 255:red, 139; green, 6; blue, 24 }  ,draw opacity=1 ][fill={rgb, 255:red, 139; green, 6; blue, 24 }  ,fill opacity=1 ] (551.23,131.53) .. controls (551.23,130.74) and (551.87,130.11) .. (552.65,130.11) .. controls (553.43,130.11) and (554.07,130.74) .. (554.07,131.53) .. controls (554.07,132.31) and (553.43,132.95) .. (552.65,132.95) .. controls (551.87,132.95) and (551.23,132.31) .. (551.23,131.53) -- cycle ;
\draw  [color={rgb, 255:red, 139; green, 6; blue, 24 }  ,draw opacity=1 ][fill={rgb, 255:red, 139; green, 6; blue, 24 }  ,fill opacity=1 ] (408.23,132.11) .. controls (408.23,131.32) and (408.87,130.68) .. (409.65,130.68) .. controls (410.43,130.68) and (411.07,131.32) .. (411.07,132.11) .. controls (411.07,132.89) and (410.43,133.53) .. (409.65,133.53) .. controls (408.87,133.53) and (408.23,132.89) .. (408.23,132.11) -- cycle ;
\draw  [color={rgb, 255:red, 139; green, 6; blue, 24 }  ,draw opacity=1 ][fill={rgb, 255:red, 139; green, 6; blue, 24 }  ,fill opacity=1 ] (270.82,132.53) .. controls (270.82,131.74) and (271.45,131.11) .. (272.23,131.11) .. controls (273.02,131.11) and (273.65,131.74) .. (273.65,132.53) .. controls (273.65,133.31) and (273.02,133.95) .. (272.23,133.95) .. controls (271.45,133.95) and (270.82,133.31) .. (270.82,132.53) -- cycle ;

\draw (109.86,75.93) node [anchor=north west][inner sep=0.75pt]    {$\CW_{3}$};
\draw (169,124.4) node [anchor=north west][inner sep=0.75pt]    {$=$};
\draw (118.65,155.35) node [anchor=north west][inner sep=0.75pt]    {$J_{1,V}$};
\draw (245.86,95.93) node [anchor=north west][inner sep=0.75pt]    {\small $\CW_{\phi(\CS)}$};
\draw (390.86,95.93) node [anchor=north west][inner sep=0.75pt]    {\small $\CW_{\phi(\CS)}$};
\draw (545.86,88.93) node [anchor=north west][inner sep=0.75pt]    {\small $\CW_{\phi(\CS)}$};
\draw (320,170.4) node [anchor=north west][inner sep=0.75pt]    {$1$};
\draw (465,172.4) node [anchor=north west][inner sep=0.75pt]    {$V_{ \begin{array}{l}
\end{array}}$};
\draw (237,223.4) node [anchor=north west][inner sep=0.75pt]    {\small $t=1$};
\draw (527,224.4) node [anchor=north west][inner sep=0.75pt]    {\small $t=3$};
\draw (385,224.4) node [anchor=north west][inner sep=0.75pt]    {\small $t=2$};
\draw (30.65,121.35) node [anchor=north west][inner sep=0.75pt]    {$L_{U_{1},U_{2}}$};

\end{tikzpicture}
\caption{A surface operators in $\CW_3$ which braids non-trivially with $L_{U_1,U_2}$ for some $U_1,U_2$ is a detectable error.}
\label{fig:detectable errors from linking}
\end{figure}

\subsection{Detectors of DSC from endable surface operators}

\begin{figure}[h!]
\centering

\tikzset{every picture/.style={line width=0.75pt}} 

\begin{tikzpicture}[x=0.75pt,y=0.75pt,yscale=-0.6,xscale=0.6]

\draw  [color={rgb, 255:red, 245; green, 166; blue, 35 }  ,draw opacity=1 ][fill={rgb, 255:red, 255; green, 224; blue, 187 }  ,fill opacity=1 ] (250.13,217) -- (250.15,118.86) -- (341.17,74.06) -- (341.15,172.19) -- cycle ;
\draw  [color={rgb, 255:red, 245; green, 166; blue, 35 }  ,draw opacity=1 ][fill={rgb, 255:red, 255; green, 224; blue, 187 }  ,fill opacity=1 ] (77.13,221) -- (77.15,122.86) -- (168.17,78.06) -- (168.15,176.19) -- cycle ;
\draw  [color={rgb, 255:red, 139; green, 6; blue, 24 }  ,draw opacity=1 ][fill={rgb, 255:red, 139; green, 6; blue, 24 }  ,fill opacity=1 ] (121.23,149.53) .. controls (121.23,148.74) and (121.87,148.11) .. (122.65,148.11) .. controls (123.43,148.11) and (124.07,148.74) .. (124.07,149.53) .. controls (124.07,150.31) and (123.43,150.95) .. (122.65,150.95) .. controls (121.87,150.95) and (121.23,150.31) .. (121.23,149.53) -- cycle ;
\draw  [color={rgb, 255:red, 139; green, 6; blue, 24 }  ,draw opacity=1 ][fill={rgb, 255:red, 139; green, 6; blue, 24 }  ,fill opacity=1 ] (298.65,145.53) .. controls (298.65,144.74) and (299.28,144.11) .. (300.07,144.11) .. controls (300.85,144.11) and (301.48,144.74) .. (301.48,145.53) .. controls (301.48,146.31) and (300.85,146.95) .. (300.07,146.95) .. controls (299.28,146.95) and (298.65,146.31) .. (298.65,145.53) -- cycle ;
\draw  [color={rgb, 255:red, 139; green, 6; blue, 24 }  ,draw opacity=1 ][fill={rgb, 255:red, 139; green, 6; blue, 24 }  ,fill opacity=1 ] (443.07,137.53) .. controls (443.07,136.74) and (443.7,136.11) .. (444.48,136.11) .. controls (445.26,136.11) and (445.9,136.74) .. (445.9,137.53) .. controls (445.9,138.31) and (445.26,138.95) .. (444.48,138.95) .. controls (443.7,138.95) and (443.07,138.31) .. (443.07,137.53) -- cycle ;
\draw [color={rgb, 255:red, 139; green, 6; blue, 24 }  ,draw opacity=1 ]   (298.65,145.53) -- (355.17,145.53) ;
\draw  [color={rgb, 255:red, 245; green, 166; blue, 35 }  ,draw opacity=1 ][fill={rgb, 255:red, 255; green, 224; blue, 187 }  ,fill opacity=1 ] (356.13,216) -- (356.15,117.86) -- (447.17,73.06) -- (447.15,171.19) -- cycle ;
\draw  [color={rgb, 255:red, 245; green, 166; blue, 35 }  ,draw opacity=1 ][fill={rgb, 255:red, 255; green, 224; blue, 187 }  ,fill opacity=1 ] (523.13,215) -- (523.15,116.86) -- (614.17,72.06) -- (614.15,170.19) -- cycle ;
\draw  [color={rgb, 255:red, 139; green, 6; blue, 24 }  ,draw opacity=1 ][fill={rgb, 255:red, 139; green, 6; blue, 24 }  ,fill opacity=1 ] (401.65,144.53) .. controls (401.65,143.74) and (402.28,143.11) .. (403.07,143.11) .. controls (403.85,143.11) and (404.48,143.74) .. (404.48,144.53) .. controls (404.48,145.31) and (403.85,145.95) .. (403.07,145.95) .. controls (402.28,145.95) and (401.65,145.31) .. (401.65,144.53) -- cycle ;
\draw  [color={rgb, 255:red, 139; green, 6; blue, 24 }  ,draw opacity=1 ][fill={rgb, 255:red, 139; green, 6; blue, 24 }  ,fill opacity=1 ] (568.65,143.53) .. controls (568.65,142.74) and (569.28,142.11) .. (570.07,142.11) .. controls (570.85,142.11) and (571.48,142.74) .. (571.48,143.53) .. controls (571.48,144.31) and (570.85,144.95) .. (570.07,144.95) .. controls (569.28,144.95) and (568.65,144.31) .. (568.65,143.53) -- cycle ;
\draw [color={rgb, 255:red, 139; green, 6; blue, 24 }  ,draw opacity=1 ]   (404.48,144.53) -- (456,144) ;
\draw [color={rgb, 255:red, 139; green, 6; blue, 24 }  ,draw opacity=1 ] [dash pattern={on 4.5pt off 4.5pt}]  (522,144) -- (569.48,143.53) ;
\draw [color={rgb, 255:red, 139; green, 6; blue, 24 }  ,draw opacity=1 ] [dash pattern={on 4.5pt off 4.5pt}]  (357.13,145.06) -- (401.65,144.53) ;
\draw [color={rgb, 255:red, 139; green, 6; blue, 24 }  ,draw opacity=1 ]   (494,144) -- (522,144) ;

\draw (121.86,103.93) node [anchor=north west][inner sep=0.75pt]    {$\CW_{\Delta}$};
\draw (202,138.4) node [anchor=north west][inner sep=0.75pt]    {$=$};
\draw (105.65,157.35) node [anchor=north west][inner sep=0.75pt]    {$L_{U_{1} ,U_{2}, \dotsc ,U_{\Delta -1}}$};
\draw (319.41,146.66) node [anchor=north west][inner sep=0.75pt]    {$U_{1}$};
\draw (422.24,146.66) node [anchor=north west][inner sep=0.75pt]    {$U_{2}$};
\draw (265.86,110.93) node [anchor=north west][inner sep=0.75pt]    {\small $\CW_{\phi(\CM_1)}$};
\draw (372.86,110.93) node [anchor=north west][inner sep=0.75pt]    {\small $\CW_{\phi(\CM_2)}$};
\draw (535.86,110.93) node [anchor=north west][inner sep=0.75pt]    {\small $\CW_{\phi(\CM_\Delta)}$};
\draw (462,139.4) node [anchor=north west][inner sep=0.75pt]    {$\dotsc $};
\draw (528.24,149.66) node [anchor=north west][inner sep=0.75pt]    {$U_{\Delta -1}$};
\draw (261,235.4) node [anchor=north west][inner sep=0.75pt]    {\small $t=1$};
\draw (380,235.4) node [anchor=north west][inner sep=0.75pt]    {\small $t=2$};
\draw (542,234.4) node [anchor=north west][inner sep=0.75pt]    {\small $t=\Delta $};

\end{tikzpicture}
\caption{The detectors of the DSC between time $t=1$ and $t=\Delta$ are obtained from endable surface operators.}
\label{fig:detectors in DSC}	
\end{figure}
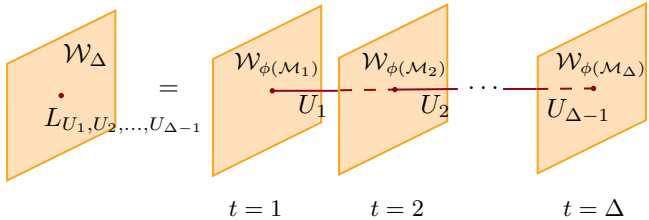

In the previous section, we saw that the detectors of a static stabilizer code defined by a stabilizer code $\CS$ correspond to surface operators in the corresponding 2-form gauge theory which can end on the 4-dimensional operator $\CW_{\phi(\CS)}$. In this section, we will generalize this to dynamic stabilizer codes. To that end, consider a sequence of measurements $\CM_1,\CM_2, \dots$ on $n$ qubits. Under the isomorphism $\phi$, this corresponds to a sequence of 4-dimensional operators $\CW_{\phi(\CM_1)},\CW_{\phi(\CM_2)},\dots$. The ISG at time $t$ is obtained by passing the trivial surface operator $\mathds{1}$ through the sequence of operators $\CW_{\phi(\CM_1)},\CW_{\phi(\CM_2)},\dots$. That is, the ISG at time $t$ is given by
\be
\CS_t = \langle \CW_{\phi(\CM_t)} \dots \CW_{\phi(\CM_2)} \circ \CW_{\phi(\CM_1)} (\mathds{1}) \rangle~,
\ee
where the expression on the right denotes the group generated by the set of surface operators produced by the composition of the action of $\CW_{\phi(\CM_i)}$ on $\mathds{1}$. Let us determine the detectors during time $t=1$ to $\Delta$. Finding the detectors and the equivalence classes of errors that can be detected in this DSC proceeds exactly as in the case of static stabilizer codes. In particular, the detectors again correspond to configurations of endable surface operators. Fig. \ref{fig:lines from endable surfaces} generalizes to Fig. \ref{fig:detectors in DSC}.
Unlike in the case of static stabilizer codes, the surface operators $U_i$ which can form such a consistent configuration, can be different for different times depending on the choice of measurements. Similar to the the case of static stabilizer codes, the detectable errors are surface operators on $\CW_{\Delta}$ which are all of the form $J_{V_1,\dots,V_{\Delta-1}}$ which braid non-trivially with the line operator $L_{U_{1},U_{2} \dots U_{\Delta -1}}$. 

\begin{figure}[h!]
\centering

\tikzset{every picture/.style={line width=0.75pt}} 

\begin{tikzpicture}[x=0.75pt,y=0.75pt,yscale=-0.65,xscale=0.65]

\draw  [color={rgb, 255:red, 245; green, 166; blue, 35 }  ,draw opacity=1 ][fill={rgb, 255:red, 255; green, 224; blue, 187 }  ,fill opacity=1 ] (96.13,217) -- (96.15,118.86) -- (187.17,74.06) -- (187.15,172.19) -- cycle ;
\draw  [color={rgb, 255:red, 139; green, 6; blue, 24 }  ,draw opacity=1 ][fill={rgb, 255:red, 139; green, 6; blue, 24 }  ,fill opacity=1 ] (144.65,131.53) .. controls (144.65,130.74) and (145.28,130.11) .. (146.07,130.11) .. controls (146.85,130.11) and (147.48,130.74) .. (147.48,131.53) .. controls (147.48,132.31) and (146.85,132.95) .. (146.07,132.95) .. controls (145.28,132.95) and (144.65,132.31) .. (144.65,131.53) -- cycle ;
\draw  [color={rgb, 255:red, 139; green, 6; blue, 24 }  ,draw opacity=1 ][fill={rgb, 255:red, 139; green, 6; blue, 24 }  ,fill opacity=1 ] (307.07,123.53) .. controls (307.07,122.74) and (307.7,122.11) .. (308.48,122.11) .. controls (309.26,122.11) and (309.9,122.74) .. (309.9,123.53) .. controls (309.9,124.31) and (309.26,124.95) .. (308.48,124.95) .. controls (307.7,124.95) and (307.07,124.31) .. (307.07,123.53) -- cycle ;
\draw [color={rgb, 255:red, 139; green, 6; blue, 24 }  ,draw opacity=1 ]   (144.65,131.53) -- (219.17,131.53) ;
\draw  [color={rgb, 255:red, 245; green, 166; blue, 35 }  ,draw opacity=1 ][fill={rgb, 255:red, 255; green, 224; blue, 187 }  ,fill opacity=1 ] (220.13,216) -- (220.15,117.86) -- (311.17,73.06) -- (311.15,171.19) -- cycle ;
\draw  [color={rgb, 255:red, 245; green, 166; blue, 35 }  ,draw opacity=1 ][fill={rgb, 255:red, 255; green, 224; blue, 187 }  ,fill opacity=1 ] (359.13,215) -- (359.15,116.86) -- (450.17,72.06) -- (450.15,170.19) -- cycle ;
\draw  [color={rgb, 255:red, 139; green, 6; blue, 24 }  ,draw opacity=1 ][fill={rgb, 255:red, 139; green, 6; blue, 24 }  ,fill opacity=1 ] (265.65,130.53) .. controls (265.65,129.74) and (266.28,129.11) .. (267.07,129.11) .. controls (267.85,129.11) and (268.48,129.74) .. (268.48,130.53) .. controls (268.48,131.31) and (267.85,131.95) .. (267.07,131.95) .. controls (266.28,131.95) and (265.65,131.31) .. (265.65,130.53) -- cycle ;
\draw  [color={rgb, 255:red, 139; green, 6; blue, 24 }  ,draw opacity=1 ][fill={rgb, 255:red, 139; green, 6; blue, 24 }  ,fill opacity=1 ] (406.82,129.53) .. controls (406.82,128.74) and (407.45,128.11) .. (408.23,128.11) .. controls (409.02,128.11) and (409.65,128.74) .. (409.65,129.53) .. controls (409.65,130.31) and (409.02,130.95) .. (408.23,130.95) .. controls (407.45,130.95) and (406.82,130.31) .. (406.82,129.53) -- cycle ;
\draw [color={rgb, 255:red, 139; green, 6; blue, 24 }  ,draw opacity=1 ]   (268.48,130.53) -- (320,130) ;
\draw [color={rgb, 255:red, 139; green, 6; blue, 24 }  ,draw opacity=1 ] [dash pattern={on 4.5pt off 4.5pt}]  (362,130) -- (409.65,129.53) ;
\draw [color={rgb, 255:red, 139; green, 6; blue, 24 }  ,draw opacity=1 ] [dash pattern={on 4.5pt off 4.5pt}]  (223.13,131.06) -- (265.65,130.53) ;
\draw  [color={rgb, 255:red, 245; green, 166; blue, 35 }  ,draw opacity=1 ][fill={rgb, 255:red, 255; green, 224; blue, 187 }  ,fill opacity=1 ] (494.13,213) -- (494.15,114.86) -- (585.17,70.06) -- (585.15,168.19) -- cycle ;
\draw  [color={rgb, 255:red, 74; green, 144; blue, 226 }  ,draw opacity=1 ][fill={rgb, 255:red, 74; green, 144; blue, 226 }  ,fill opacity=1 ] (538.07,165.95) .. controls (538.07,165.16) and (538.7,164.53) .. (539.48,164.53) .. controls (540.26,164.53) and (540.9,165.16) .. (540.9,165.95) .. controls (540.9,166.74) and (540.26,167.37) .. (539.48,167.37) .. controls (538.7,167.37) and (538.07,166.74) .. (538.07,165.95) -- cycle ;
\draw [color={rgb, 255:red, 74; green, 144; blue, 226 }  ,draw opacity=1 ] [dash pattern={on 4.5pt off 4.5pt}]  (497.42,166.42) -- (540.9,165.95) ;
\draw [color={rgb, 255:red, 74; green, 144; blue, 226 }  ,draw opacity=1 ]   (404.42,165.58) -- (494,166) ;
\draw [color={rgb, 255:red, 139; green, 6; blue, 24 }  ,draw opacity=1 ]   (320,130) -- (360,130) ;
\draw  [color={rgb, 255:red, 74; green, 144; blue, 226 }  ,draw opacity=1 ][fill={rgb, 255:red, 74; green, 144; blue, 226 }  ,fill opacity=1 ] (401.58,165.58) .. controls (401.58,164.79) and (402.22,164.15) .. (403,164.15) .. controls (403.78,164.15) and (404.42,164.79) .. (404.42,165.58) .. controls (404.42,166.36) and (403.78,167) .. (403,167) .. controls (402.22,167) and (401.58,166.36) .. (401.58,165.58) -- cycle ;
\draw [color={rgb, 255:red, 74; green, 144; blue, 226 }  ,draw opacity=1 ] [dash pattern={on 4.5pt off 4.5pt}]  (362.1,166.05) -- (402.62,165.58) -- (403,165.58) ;
\draw [color={rgb, 255:red, 74; green, 144; blue, 226 }  ,draw opacity=1 ]   (269,166) -- (360.1,166.05) ;
\draw  [color={rgb, 255:red, 74; green, 144; blue, 226 }  ,draw opacity=1 ][fill={rgb, 255:red, 74; green, 144; blue, 226 }  ,fill opacity=1 ] (265.17,166) .. controls (265.17,165.21) and (265.8,164.58) .. (266.58,164.58) .. controls (267.37,164.58) and (268,165.21) .. (268,166) .. controls (268,166.79) and (267.37,167.42) .. (266.58,167.42) .. controls (265.8,167.42) and (265.17,166.79) .. (265.17,166) -- cycle ;

\draw (154.91,134.93) node [anchor=north west][inner sep=0.75pt]    {\small $X_{1} X_{2} X_{3} X_{4}$};
\draw (140.86,95.93) node [anchor=north west][inner sep=0.75pt]    {\small $\CW_{\CM_2}$};
\draw (263.86,95.93) node [anchor=north west][inner sep=0.75pt]    {\small $\CW_{\CM_1}$};
\draw (401.86,95.93) node [anchor=north west][inner sep=0.75pt]    {\small $\CW_{\CM_1}$};
\draw (119,225.4) node [anchor=north west][inner sep=0.75pt]    {\small $t=2$};
\draw (246,227.4) node [anchor=north west][inner sep=0.75pt]    {\small $t=3$};
\draw (381,225.4) node [anchor=north west][inner sep=0.75pt]    {\small $t=4$};
\draw (537.86,93.93) node [anchor=north west][inner sep=0.75pt]    {\small $\CW_{\CM_2}$};
\draw (514,225.4) node [anchor=north west][inner sep=0.75pt]    {\small $t=5$};
\draw (290.41,133.66) node [anchor=north west][inner sep=0.75pt]    {\small $X_{1} X_{2} X_{3} X_{4}$};
\draw (280.91,167.93) node [anchor=north west][inner sep=0.75pt]    {\small $Z_{1} Z_{2} Z_{3} Z_{4}$};
\draw (431.21,170.19) node [anchor=north west][inner sep=0.75pt]    {\small $Z_{1} Z_{2} Z_{3} Z_{4}$};

\end{tikzpicture}
\caption{In the Bacon-Shor DSC, there are two endable surface operators which generate the detectors of the code. These correspond to the $X_1X_2X_3X_4$ and $Z_1Z_2Z_3Z_4$ Pauli matrices.}
\label{fig: example detectors}
\end{figure}
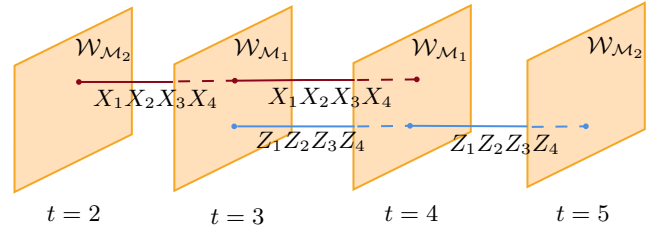

Let us consider the Bacon-Shor DSC example again with $\CM_1=\langle Z_1Z_2, Z_3Z_4\rangle, \CM_{2}=\langle X_1X_3,X_2X_4\rangle$ repeated indefinitely. We will consider this DSC during the time $t=2$ to $t=5$. The detectors of the code are given by endable surface operators. These are determined in Fig. \ref{fig: example detectors} which agree with the detectors found in \eqref{eq:Bacon-Shor detectors}.

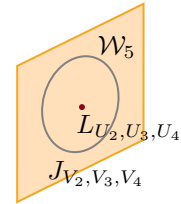
\begin{figure}[h!]
    \centering

\tikzset{every picture/.style={line width=0.75pt}} 

\begin{tikzpicture}[x=0.75pt,y=0.75pt,yscale=-0.7,xscale=0.7]

\draw  [color={rgb, 255:red, 245; green, 166; blue, 35 }  ,draw opacity=1 ][fill={rgb, 255:red, 255; green, 224; blue, 187 }  ,fill opacity=1 ] (280.98,215.97) -- (281,117.83) -- (372.02,73.03) -- (372,171.17) -- cycle ;
\draw  [color={rgb, 255:red, 128; green, 128; blue, 128 }  ,draw opacity=1 ] (299.81,158.78) .. controls (296.02,141.16) and (304.89,120.92) .. (319.62,113.56) .. controls (334.35,106.19) and (349.36,114.5) .. (353.15,132.11) .. controls (356.93,149.72) and (348.06,169.97) .. (333.33,177.33) .. controls (318.6,184.7) and (303.59,176.39) .. (299.81,158.78) -- cycle ;
\draw  [color={rgb, 255:red, 139; green, 6; blue, 24 }  ,draw opacity=1 ][fill={rgb, 255:red, 139; green, 6; blue, 24 }  ,fill opacity=1 ] (326.23,147.53) .. controls (326.23,146.74) and (326.87,146.11) .. (327.65,146.11) .. controls (328.43,146.11) and (329.07,146.74) .. (329.07,147.53) .. controls (329.07,148.31) and (328.43,148.95) .. (327.65,148.95) .. controls (326.87,148.95) and (326.23,148.31) .. (326.23,147.53) -- cycle ;

\draw (336.86,93.93) node [anchor=north west][inner sep=0.75pt]    {$\CW_{5}$};
\draw (300.65,184.35) node [anchor=north west][inner sep=0.75pt]    {$J_{V_{2} ,V_{3} ,V_{4}}$};
\draw (321.65,151.35) node [anchor=north west][inner sep=0.75pt]    {$L_{U_{2} ,U_{3} ,U_{4}}$};

\end{tikzpicture}
    \caption{Fusing the four non-invertible symmetry operators $\CW_{\CM_i}$ in Fig. \ref{fig: example detectors} gives an operator $\CW_{5}$. The detectable errors of the Bacon-Shor Dynamical Code are determined by the braiding of line and surface operators on $\CW_{5}$.}
    \label{fig:detectable errors example}
\end{figure}

\noindent On fusing the four $\CW_{\CM_i}$ operators, we get the operator $\CW_5$ with line operators $L_{U_2,U_3,U_4}$. From the detectors found in Fig. \ref{fig: example detectors} we find that the line operators $L_{U_2,U_3,U_4}$ on $\CW_5$ are generated by $L_{X_1X_2X_3X_4,X_1X_2X_3X_4,\mathds{1}}$ and $L_{\mathds{1},Z_1Z_2Z_3Z_4,Z_1Z_2Z_3Z_4}$. A general surface operator on $\CW_{5}$ is of the form $J_{V_2,V_3,V_4}$ where $V_i$ is a surface operator acting after time $t=i$. Under the isomorphism $\phi$, $V_2,\dots,V_4$ is a set of errors that can occur during the time $t=2$ to $t=5$. This error can be detected if the surface operator $J_{V_2,V_3,V_4}$ braids non-trivially with the at least one of the line operators $L_{X_1X_2X_3X_4,X_1X_2X_3X_4,\mathds{1}}$ and $L_{\mathds{1},Z_1Z_2Z_3Z_4,Z_1Z_2Z_3Z_4}$ (see Fig. \ref{fig:detectable errors example}).

\subsection{Relation to spacetime stabilizer code}

In \cite{Delfosse:2023aqk}, the authors show that the detectors of a DSC can be used to construct a static stabilizer code called the \textit{spacetime} code. In this way, decoding DSC is mapped to a familiar problem of decoding in a static stabilizer code. To understand their framework, consider an error $E_{i,t}$, where $1\leq i\leq n$ denotes a qubit and $t$ denotes time. $E_{i,t}$ is an error acting on the $i^{\rm th}$ qubit right before the timestep $t$. Therefore, a general error that can occur at time $t$ is an $n$-qubit Pauli operator of the form
\be
\prod_{i=1}^{n} E_{i,t} ~.
\ee
Consider a finite number of time steps $t=1,\dots \Delta$. A general Pauli error can be denoted as
\be
\label{eq:spacetime error}
E=\bigotimes_{t=1}^{\Delta -1} \prod_{i=1}^{n} E_{i,t}~.
\ee
This operator can be regarded as a Pauli matrix acting on $n(\Delta -1)$ qubits. Consider the error $E_{i,t}$ acting on the $i^{\rm th}$ qubit before timestep $t$. Suppose $M_t \in \CM_t$ and let $M_{i,t}$ be the support of the Pauli matrix $M_t$ on the $i^{\rm th}$ qubit. If $E_{i,t}$ anti-commutes with $M_{i,t}$, then the measurement outcome $O(M_{i,t})$ is flipped by this error. The cumulative effect of the error $E$ on the measurement outcome $O(M_{i,t})$ is determined by not just $E_{i,t}$, but rather the combination of all errors that occurred on the qubit $i$ before time $t$. Therefore the relevant commutation relation is $[M_{i,t},F(E)_{i,t}]$, where $F(E)$ is an $n(\Delta-1)$ qubit Pauli operators called the \textit{cumulant} of the error $E$ whose action on the $i^{\rm th}$ qubit before time $t$ is defined as 
\be
\label{eq:cumulant}
F(E)_{i,t}:= \prod_{1 \leq t'<t} E_{i,t'}~. 
\ee
Therefore, the error $E$ can be detected if the cumulant $F(E)$ violates at least one of the detectors $\CD$. In \cite{Delfosse:2023aqk}, the authors show that 
\be
[F(E),D]=[E,K(D)]~.
\ee
where $K(D)$ is called the \textit{back-cumulant} of a detector. It is shown that $K(D)$ for $D \in \CD$ are generators of a static stabilizer code acting on $n(\Delta - 1)$ qubits.

In our framework, the cumulant of an error is passing an error through the sequence of operators $\CW_{\CM_i}$ and the back-cumulant of a detector is precisely the endable surface operators as defined in Fig. \ref{fig:detectors in DSC}. The topological fact that line operators $L_{U_1,\dots,U_{\Delta-1}}$ on the operator $\CW_{\Delta}$ braid trivially with each other is consistent with the fact that the back-cumulant of detectors defines a stabilizer code.

\section{Conclusion}

We showed that static stabilizer codes on $n$-qudits are in one-to-one correspondence with non-anomalous symmetries of a 2-form gauge theory. Moreover, a sequence of measurements corresponds to a sequence of non-invertible symmetries implemented by 4-dimensional operators in the gauge theory. We use this framework to study a dynamical stabilizer code using a 2-form gauge theory. In particular, we showed that the detectors of the DSC are given by endable surface operators. In fact, we can choose more general $K_{ij}$ in the action \eqref{eq:2-form gauge theory action} to get DSCs defined on qubits or a system of qudits of various dimensions. Therefore, 2-form gauge theories provide a unified framework to study general DSCs in composite qudit dimensions. Moreover, by inserting invertible 4-dimensional operators between the non-invertible operators, our setting also generalizes to shallow depth circuits. In this setting, several natural questions arise:
\begin{itemize}
\item General high-weight Pauli measurements can be reduced to a sequence of non-commuting low-weight Pauli measurements. Using ZX calculus, this can be done in such a way that the resulting map from a static stabilizer code to a DSC is distance-preserving \cite{Townsend-Teague23,Rodatz:2024jkm}. It will be good to understand if the gauge theory description of both static and dynamical codes gives a complimentary method to do the same.
	\item Do good quantum LDPC codes and their floquetification have a nice interpreration in terms of the 2-form gauge theory? In fact, good static quantum LDPC codes require all-to-all connectivity between qubits for implementing the measurements in the stabilizer group. A natural question is whether a dynamical version of such codes can help in reducing the connectivity required and whether this problem translates to a natural question in the gauge theory. 
	\item In this paper, we only considered 2-form gauge theories, which are special 4+1d TQFTs whose two dimensional surface operators form a group. More generally, we can consider 4+1d TQFTs in which the surface operators are non-invertible \cite{Johnson-Freyd:2021tbq}. This opens the door to static quantum error correcting codes where the logical information is protected from errors by non-invertible symmetries.  
	\item Given a DSC, it is important to understand the logical gates that can be fault-tolerantly implemented on the code subspace. These must correspond to various 0-form symmetries of the 2-form gauge theory. Since the 2-form gauge theory is a TQFT, its non-trivial 0-form symmetries are all implemented by 4-dimensional operators obtained from higher-gauging the surface operators. It will be interesting to understand the consequences of this construction for the corresponding DSC. 
\end{itemize}
We hope to return to these questions in the future. 
\\

\noindent Note: During the write-up stage of this work, the paper \cite{Barbar:2025krh} appeared, which has minor overlap with our work. 

\section{Acknowledgments} 

RR would like to thank Matthew Buican, Clement Delcamp, and Anatoly Dymarsky for discussions related to this project. This work was initiated during a visit to the Simons Center for Geometry and Physics, whose hospitality is gratefully acknowledged. RR also thanks the International Centre for Mathematical Sciences for their hospitality during a Research-in-Groups programme. The work of RR is supported by the UKRI Frontier Research Grant, underwriting the ERC Advanced Grant ``Generalized Symmetries in Quantum Field Theory and Quantum Gravity".

\bibliographystyle{ytphys}
\bibliography{ref}

\providecommand{\href}[2]{#2}\begingroup\raggedright\begin{thebibliography}{10}

\bibitem{Calderbank97}
A.~R. Calderbank, E.~M. Rains, P.~W. Shor, and N.~J.~A. Sloane, {\slshape Quantum error correction and orthogonal geometry,} \href{https://link.aps.org/doi/10.1103/PhysRevLett.78.405}{{\em Phys. Rev. Lett.} {\bfseries 78} (Jan, 1997) 405--408}.

\bibitem{Gottesman:1997zz}
D.~Gottesman, {\slshape {Stabilizer codes and quantum error correction},} \href{http://arxiv.org/abs/quant-ph/9705052}{{ arXiv:quant-ph/9705052}}.

\bibitem{Gottesman:1998se}
D.~Gottesman, {\slshape {Fault tolerant quantum computation with higher dimensional systems},} \href{http://dx.doi.org/10.1016/S0960-0779(98)00218-5}{{\em Chaos Solitons Fractals} {\bfseries 10} (1999) 1749--1758}, \href{http://arxiv.org/abs/quant-ph/9802007}{{ arXiv:quant-ph/9802007}}.

\bibitem{hastings2021dynamically}
M.~B. Hastings and J.~Haah, {\slshape Dynamically {G}enerated {L}ogical {Q}ubits,} \href{https://doi.org/10.22331/q-2021-10-19-564}{{\em {Quantum}} {\bfseries 5} (Oct., 2021) 564}.

\bibitem{Haah2022boundarieshoneycomb}
J.~Haah and M.~B. Hastings, {\slshape Boundaries for the {H}oneycomb {C}ode,} \href{https://doi.org/10.22331/q-2022-04-21-693}{{\em {Quantum}} {\bfseries 6} (Apr., 2022) 693}.

\bibitem{Aasen22}
D.~Aasen, Z.~Wang, and M.~B. Hastings, {\slshape Adiabatic paths of hamiltonians, symmetries of topological order, and automorphism codes,} \href{https://link.aps.org/doi/10.1103/PhysRevB.106.085122}{{\em Phys. Rev. B} {\bfseries 106} (Aug, 2022) 085122}.

\bibitem{Gidney2021faulttolerant}
C.~Gidney, M.~Newman, A.~Fowler, and M.~Broughton, {\slshape A {F}ault-{T}olerant {H}oneycomb {M}emory,} \href{https://doi.org/10.22331/q-2021-12-20-605}{{\em {Quantum}} {\bfseries 5} (Dec., 2021) 605}.

\bibitem{Davydova23}
M.~Davydova, N.~Tantivasadakarn, and S.~Balasubramanian, {\slshape Floquet codes without parent subsystem codes,} \href{https://link.aps.org/doi/10.1103/PRXQuantum.4.020341}{{\em PRX Quantum} {\bfseries 4} (Jun, 2023) 020341}.

\bibitem{Kesselring:2022eax}
M.~S. Kesselring, J.~C.~M. de~la Fuente, F.~Thomsen, J.~Eisert, S.~D. Bartlett, and B.~J. Brown, {\slshape {Anyon Condensation and the Color Code},} \href{http://dx.doi.org/10.1103/PRXQuantum.5.010342}{{\em PRX Quantum} {\bfseries 5} (2024) 010342}, \href{http://arxiv.org/abs/2212.00042}{{ arXiv:2212.00042~[quant-ph]}}.

\bibitem{Zhang23}
Z.~Zhang, D.~Aasen, and S.~Vijay, {\slshape $x$-cube floquet code: A dynamical quantum error correcting code with a subextensive number of logical qubits,} \href{https://link.aps.org/doi/10.1103/PhysRevB.108.205116}{{\em Phys. Rev. B} {\bfseries 108} (Nov, 2023) 205116}.

\bibitem{Bombin2024}
H.~Bombin, D.~Litinski, N.~Nickerson, F.~Pastawski, and S.~Roberts, {\slshape Unifying flavors of fault tolerance with the {ZX} calculus,} \href{https://doi.org/10.22331/q-2024-06-18-1379}{{\em {Quantum}} {\bfseries 8} (June, 2024) 1379}.

\bibitem{Gidney2023}
C.~Gidney, {\slshape A {P}air {M}easurement {S}urface {C}ode on {P}entagons,} \href{https://doi.org/10.22331/q-2023-10-25-1156}{{\em {Quantum}} {\bfseries 7} (Oct., 2023) 1156}.

\bibitem{Bauer2024topologicalerror}
A.~Bauer, {\slshape Topological error correcting processes from fixed-point path integrals,} \href{https://doi.org/10.22331/q-2024-03-20-1288}{{\em {Quantum}} {\bfseries 8} (Mar., 2024) 1288}.

\bibitem{Ellison23}
T.~D. Ellison, J.~Sullivan, and A.~Dua, {\slshape Floquet codes with a twist,} {\em arXiv preprint arXiv:2306.08027} (2023) .

\bibitem{Davydova:2023mnz}
M.~Davydova, N.~Tantivasadakarn, S.~Balasubramanian, and D.~Aasen, {\slshape {Quantum computation from dynamic automorphism codes},} \href{http://dx.doi.org/10.22331/q-2024-08-27-1448}{{\em Quantum} {\bfseries 8} (2024) 1448}, \href{http://arxiv.org/abs/2307.10353}{{ arXiv:2307.10353~[quant-ph]}}.

\bibitem{Dua24}
A.~Dua, N.~Tantivasadakarn, J.~Sullivan, and T.~D. Ellison, {\slshape Engineering 3d floquet codes by rewinding,} \href{https://link.aps.org/doi/10.1103/PRXQuantum.5.020305}{{\em PRX Quantum} {\bfseries 5} (Apr, 2024) 020305}.

\bibitem{Townsend-Teague23}
A.~Townsend-Teague, J.~M. de~la Fuente, and M.~Kesselring, {\slshape {Floquetifying the Colour Code},} \href{http://dx.doi.org/10.4204/EPTCS.384.14}{{\em EPTCS} {\bfseries 384} (2023) 265--303}, \href{http://arxiv.org/abs/2307.11136}{{ arXiv:2307.11136~[quant-ph]}}.

\bibitem{Kobayashi:2023zxs}
R.~Kobayashi and G.~Zhu, {\slshape {Cross-Cap Defects and Fault-Tolerant Logical Gates in the Surface Code and the Honeycomb Floquet Code},} \href{http://dx.doi.org/10.1103/PRXQuantum.5.020360}{{\em PRX Quantum} {\bfseries 5} (2024) 020360}, \href{http://arxiv.org/abs/2310.06917}{{ arXiv:2310.06917~[quant-ph]}}.

\bibitem{Higgott24}
O.~Higgott and N.~P. Breuckmann, {\slshape Constructions and performance of hyperbolic and semi-hyperbolic floquet codes,} \href{https://link.aps.org/doi/10.1103/PRXQuantum.5.040327}{{\em PRX Quantum} {\bfseries 5} (Nov, 2024) 040327}.

\bibitem{GransSamuelsson2024improvedpairwise}
L.~Grans-Samuelsson, R.~V. Mishmash, D.~Aasen, C.~Knapp, B.~Bauer, B.~Lackey, M.~P.~d. Silva, and P.~Bonderson, {\slshape Improved {P}airwise {M}easurement-{B}ased {S}urface {C}ode,} \href{https://doi.org/10.22331/q-2024-08-02-1429}{{\em {Quantum}} {\bfseries 8} (Aug., 2024) 1429}.

\bibitem{Fahimniya25}
A.~Fahimniya, H.~Dehghani, K.~Bharti, S.~Mathew, A.~J. Koll{\'{a}}r, A.~V. Gorshkov, and M.~J. Gullans, {\slshape Fault-tolerant hyperbolic {F}loquet quantum error correcting codes,} \href{https://doi.org/10.22331/q-2025-09-05-1849}{{\em {Quantum}} {\bfseries 9} (Sept., 2025) 1849}.

\bibitem{Bauer2025lowoverheadnon}
A.~Bauer, {\slshape Low-overhead non-{C}lifford fault-tolerant circuits for all non-chiral abelian topological phases,} \href{https://doi.org/10.22331/q-2025-03-25-1673}{{\em {Quantum}} {\bfseries 9} (Mar., 2025) 1673}.

\bibitem{Bauer2025x+y}
A.~Bauer, {\slshape $x+y$ floquet code: A simple example for topological quantum computation in the path-integral approach,} \href{https://link.aps.org/doi/10.1103/PhysRevA.111.032413}{{\em Phys. Rev. A} {\bfseries 111} (Mar, 2025) 032413}.

\bibitem{MdlF25}
J.~C. Magdalena de~la Fuente, J.~Old, A.~Townsend-Teague, M.~Rispler, J.~Eisert, and M.~M\"uller, {\slshape $\mathrm{XYZ}$ ruby code: Making a case for a three-colored graphical calculus for quantum error correction in spacetime,} \href{https://link.aps.org/doi/10.1103/PRXQuantum.6.010360}{{\em PRX Quantum} {\bfseries 6} (Mar, 2025) 010360}.

\bibitem{Xu:2025zos}
Y.~Xu and A.~Dua, {\slshape {Fault-tolerant protocols through spacetime concatenation},}
\newblock 4, 2025.
\newblock \href{http://arxiv.org/abs/2504.08918}{{ arXiv:2504.08918~[quant-ph]}}.

\bibitem{Delfosse:2023aqk}
N.~Delfosse and A.~Paetznick, {\slshape {Spacetime codes of Clifford circuits},} \href{http://arxiv.org/abs/2304.05943}{{ arXiv:2304.05943~[quant-ph]}}.

\bibitem{Fu:2024hin}
X.~Fu and D.~Gottesman, {\slshape Error correction in dynamical codes,} {\em arXiv preprint arXiv:2403.04163} (2024) .

\bibitem{Blackwell25}
K.~Blackwell and J.~Haah, {\slshape {The code distance of Floquet codes},} \href{http://arxiv.org/abs/2510.05549}{{ arXiv:2510.05549~[quant-ph]}}.

\bibitem{Kitaev:1997wr}
A.~Y. Kitaev, {\slshape {Fault tolerant quantum computation by anyons},} \href{http://dx.doi.org/10.1016/S0003-4916(02)00018-0}{{\em Annals Phys.} {\bfseries 303} (2003) 2--30}, \href{http://arxiv.org/abs/quant-ph/9707021}{{ arXiv:quant-ph/9707021}}.

\bibitem{Bombin:2013osf}
H.~Bombin, {\slshape {An Introduction to Topological Quantum Codes},} \href{http://arxiv.org/abs/1311.0277}{{ arXiv:1311.0277~[quant-ph]}}.

\bibitem{tillich2013quantum}
J.-P. Tillich and G.~Z{\'e}mor, {\slshape Quantum ldpc codes with positive rate and minimum distance proportional to the square root of the blocklength,} {\em IEEE Transactions on Information Theory} {\bfseries 60} (2013) 1193--1202.

\bibitem{bravyi2014homological}
S.~Bravyi and M.~B. Hastings, {\slshape Homological product codes,} in {\em Proceedings of the forty-sixth annual ACM symposium on Theory of computing}, pp.~273--282.
\newblock 2014.

\bibitem{hastings2021fiber}
M.~B. Hastings, J.~Haah, and R.~O'Donnell, {\slshape Fiber bundle codes: breaking the n 1/2 polylog (n) barrier for quantum ldpc codes,} in {\em Proceedings of the 53rd Annual ACM SIGACT Symposium on Theory of Computing}, pp.~1276--1288.
\newblock 2021.

\bibitem{Panteleev:2021wvc}
P.~Panteleev and G.~Kalachev, \href{http://dx.doi.org/10.1145/3519935.3520017}{{\slshape {Asymptotically good Quantum and locally testable classical LDPC codes},}} in {\em {54th Annual ACM Symposium on Theory of Computing}}.
\newblock 11, 2021.
\newblock \href{http://arxiv.org/abs/2111.03654}{{ arXiv:2111.03654~[cs.IT]}}.

\bibitem{Breuckmann:2020jpn}
N.~P. Breuckmann and J.~N. Eberhardt, {\slshape {Balanced Product Quantum Codes},} \href{http://dx.doi.org/10.1109/TIT.2021.3097347}{{\em IEEE Trans. Info. Theor.} {\bfseries 67} (2021) 6653--6674}, \href{http://arxiv.org/abs/2012.09271}{{ arXiv:2012.09271~[quant-ph]}}.

\bibitem{bravyi2009no}
S.~Bravyi and B.~Terhal, {\slshape A no-go theorem for a two-dimensional self-correcting quantum memory based on stabilizer codes,} {\em New Journal of Physics} {\bfseries 11} (2009) 043029.

\bibitem{bravyi2010tradeoffs}
S.~Bravyi, D.~Poulin, and B.~Terhal, {\slshape Tradeoffs for reliable quantum information storage in 2d systems,} {\em Physical review letters} {\bfseries 104} (2010) 050503.

\bibitem{Yin:2024hbx}
C.~Yin and A.~Lucas, {\slshape {Low-Density Parity-Check Codes as Stable Phases of Quantum Matter},} \href{http://dx.doi.org/10.1103/361k-nj4b}{{\em PRX Quantum} {\bfseries 6} (2025) 030329}, \href{http://arxiv.org/abs/2411.01002}{{ arXiv:2411.01002~[quant-ph]}}.

\bibitem{DeRoeck:2024alc}
W.~De~Roeck, V.~Khemani, Y.~Li, N.~O'Dea, and T.~Rakovszky, {\slshape {Low-Density Parity-Check Stabilizer Codes as Gapped Quantum Phases: Stability under Graph-Local Perturbations},} \href{http://dx.doi.org/10.1103/7x71-8j7k}{{\em PRX Quantum} {\bfseries 6} (2025) 030330}, \href{http://arxiv.org/abs/2411.02384}{{ arXiv:2411.02384~[quant-ph]}}.

\bibitem{Note1}
The primary operators of certain 1+1d RCFTs and the corresponding line operators of their associated 3d Chern-Simons theories capture the properties of some static qudit stabilizer codes. However, this setting is limited to special stabilizer codes. This boils down to the fact that the braiding of line operators in 2+1d is symmetric, while the abelianized Pauli group has a symplectic structure \protect \cite {Dymarsky:2020qom,Dymarsky:2021xfc,Buican:2021uyp,Buican:2023ehi}.

\bibitem{farinholt2014ideal}
J.~Farinholt, {\slshape An ideal characterization of the clifford operators,} {\em Journal of Physics A: Mathematical and Theoretical} {\bfseries 47} (2014) 305303.

\bibitem{haah2016algebraic}
J.~Haah, {\slshape Algebraic methods for quantum codes on lattices,} {\em arXiv preprint arXiv:1607.01387} (2016) .

\bibitem{Aasen16}
D.~Aasen, M.~Hell, R.~V. Mishmash, A.~Higginbotham, J.~Danon, M.~Leijnse, T.~S. Jespersen, J.~A. Folk, C.~M. Marcus, K.~Flensberg, and J.~Alicea, {\slshape Milestones toward majorana-based quantum computing,} \href{https://link.aps.org/doi/10.1103/PhysRevX.6.031016}{{\em Phys. Rev. X} {\bfseries 6} (Aug, 2016) 031016}.

\bibitem{Bartolucci2023fusion}
S.~Bartolucci, P.~Birchall, H.~Bomb{\'\i}n, H.~Cable, C.~Dawson, M.~Gimeno-Segovia, E.~Johnston, K.~Kieling, N.~Nickerson, M.~Pant, F.~Pastawski, T.~Rudolph, and C.~Sparrow, {\slshape Fusion-based quantum computation,} \href{https://doi.org/10.1038/s41467-023-36493-1}{{\em Nature Communications} {\bfseries 14} (2023) 912}.

\bibitem{delaFuente:2024ttz}
J.~C.~M. de~la Fuente, {\slshape {Dynamical weight reduction of Pauli measurements},} \href{http://arxiv.org/abs/2410.12527}{{ arXiv:2410.12527~[quant-ph]}}.

\bibitem{Chao20}
R.~Chao, M.~E. Beverland, N.~Delfosse, and J.~Haah, {\slshape Optimization of the surface code design for {M}ajorana-based qubits,} \href{https://doi.org/10.22331/q-2020-10-28-352}{{\em {Quantum}} {\bfseries 4} (Oct., 2020) 352}.

\bibitem{Aasen2023pjk}
D.~Aasen, J.~Haah, Z.~Li, and R.~S.~K. Mong, {\slshape {Measurement Quantum Cellular Automata and Anomalies in Floquet Codes},} \href{http://arxiv.org/abs/2304.01277}{{ arXiv:2304.01277~[quant-ph]}}.

\bibitem{Banks:2010zn}
T.~Banks and N.~Seiberg, {\slshape {Symmetries and Strings in Field Theory and Gravity},} \href{http://dx.doi.org/10.1103/PhysRevD.83.084019}{{\em Phys. Rev. D} {\bfseries 83} (2011) 084019}, \href{http://arxiv.org/abs/1011.5120}{{ arXiv:1011.5120~[hep-th]}}.

\bibitem{Kapustin:2014gua}
A.~Kapustin and N.~Seiberg, {\slshape {Coupling a QFT to a TQFT and Duality},} \href{http://dx.doi.org/10.1007/JHEP04(2014)001}{{\em JHEP} {\bfseries 04} (2014) 001}, \href{http://arxiv.org/abs/1401.0740}{{ arXiv:1401.0740~[hep-th]}}.

\bibitem{Chen:2021xuc}
X.~Chen, A.~Dua, P.-S. Hsin, C.-M. Jian, W.~Shirley, and C.~Xu, {\slshape {Loops in 4+1d topological phases},} \href{http://dx.doi.org/10.21468/SciPostPhys.15.1.001}{{\em SciPost Phys.} {\bfseries 15} (2023) 001}, \href{http://arxiv.org/abs/2112.02137}{{ arXiv:2112.02137~[cond-mat.str-el]}}.

\bibitem{Buican:2023bzl}
M.~Buican and R.~Radhakrishnan, {\slshape {Invertibility of Condensation Defects and Symmetries of 2 + 1d QFTs},} \href{http://dx.doi.org/10.1007/s00220-024-05096-2}{{\em Commun. Math. Phys.} {\bfseries 405} (2024) 217}, \href{http://arxiv.org/abs/2309.15181}{{ arXiv:2309.15181~[hep-th]}}.

\bibitem{KNBalasubramanian:2025vum}
M.~K.~N.~Balasubramanian, M.~Buican, C.~Delcamp, and R.~Radhakrishnan, {\slshape {Gauging Non-Invertible Symmetries in (2+1)d Topological Orders},} \href{http://arxiv.org/abs/2507.01142}{{ arXiv:2507.01142~[hep-th]}}.

\bibitem{Roumpedakis:2022aik}
K.~Roumpedakis, S.~Seifnashri, and S.-H. Shao, {\slshape {Higher Gauging and Non-invertible Condensation Defects},} \href{http://dx.doi.org/10.1007/s00220-023-04706-9}{{\em Commun. Math. Phys.} {\bfseries 401} (2023) 3043--3107}, \href{http://arxiv.org/abs/2204.02407}{{ arXiv:2204.02407~[hep-th]}}.

\bibitem{davydov2007twisted}
A.~Davydov, {\slshape Twisted automorphisms of hopf algebras,} {\em arXiv preprint arXiv:0708.2757} (2007) .

\bibitem{Johnson-Freyd:2021tbq}
T.~Johnson-Freyd and M.~Yu, {\slshape {Topological Orders in (4+1)-Dimensions},} \href{http://dx.doi.org/10.21468/SciPostPhys.13.3.068}{{\em SciPost Phys.} {\bfseries 13} (2022) 068}, \href{http://arxiv.org/abs/2104.04534}{{ arXiv:2104.04534~[hep-th]}}.

\bibitem{Gaiotto:2020iye}
D.~Gaiotto and J.~Kulp, {\slshape {Orbifold groupoids},} \href{http://dx.doi.org/10.1007/JHEP02(2021)132}{{\em JHEP} {\bfseries 02} (2021) 132}, \href{http://arxiv.org/abs/2008.05960}{{ arXiv:2008.05960~[hep-th]}}.

\bibitem{Apruzzi:2021nmk}
F.~Apruzzi, F.~Bonetti, I.~n. Garc\'\i{}a~Etxebarria, S.~S. Hosseini, and S.~Schafer-Nameki, {\slshape {Symmetry TFTs from String Theory},} \href{http://dx.doi.org/10.1007/s00220-023-04737-2}{{\em Commun. Math. Phys.} {\bfseries 402} (2023) 895--949}, \href{http://arxiv.org/abs/2112.02092}{{ arXiv:2112.02092~[hep-th]}}.

\bibitem{Chatterjee:2022kxb}
A.~Chatterjee and X.-G. Wen, {\slshape {Symmetry as a shadow of topological order and a derivation of topological holographic principle},} \href{http://dx.doi.org/10.1103/PhysRevB.107.155136}{{\em Phys. Rev. B} {\bfseries 107} (2023) 155136}, \href{http://arxiv.org/abs/2203.03596}{{ arXiv:2203.03596~[cond-mat.str-el]}}.

\bibitem{Freed:2022qnc}
D.~S. Freed, G.~W. Moore, and C.~Teleman, {\slshape {Topological symmetry in quantum field theory},} \href{http://arxiv.org/abs/2209.07471}{{ arXiv:2209.07471~[hep-th]}}.

\bibitem{Kaidi:2022cpf}
J.~Kaidi, K.~Ohmori, and Y.~Zheng, {\slshape {Symmetry TFTs for Non-invertible Defects},} \href{http://dx.doi.org/10.1007/s00220-023-04859-7}{{\em Commun. Math. Phys.} {\bfseries 404} (2023) 1021--1124}, \href{http://arxiv.org/abs/2209.11062}{{ arXiv:2209.11062~[hep-th]}}.

\bibitem{Rodatz:2024jkm}
B.~Rodatz, B.~Po{\'o}r, and A.~Kissinger, {\slshape {Floquetifying stabiliser codes with distance-preserving rewrites},} \href{http://arxiv.org/abs/2410.17240}{{ arXiv:2410.17240~[quant-ph]}}.

\bibitem{Barbar:2025krh}
A.~Barbar, A.~Dymarsky, and A.~Shapere, {\slshape {Holographic description of 4d Maxwell theories and their code-based ensembles},} \href{http://arxiv.org/abs/2510.03392}{{ arXiv:2510.03392~[hep-th]}}.

\bibitem{Dymarsky:2020qom}
A.~Dymarsky and A.~Shapere, {\slshape {Quantum stabilizer codes, lattices, and CFTs},} \href{http://dx.doi.org/10.1007/JHEP03(2021)160}{{\em JHEP} {\bfseries 21} (2020) 160}, \href{http://arxiv.org/abs/2009.01244}{{ arXiv:2009.01244~[hep-th]}}.

\bibitem{Dymarsky:2021xfc}
A.~Dymarsky and A.~Sharon, {\slshape {Non-rational Narain CFTs from codes over F$_{4}$},} \href{http://dx.doi.org/10.1007/JHEP11(2021)016}{{\em JHEP} {\bfseries 11} (2021) 016}, \href{http://arxiv.org/abs/2107.02816}{{ arXiv:2107.02816~[hep-th]}}.

\bibitem{Buican:2021uyp}
M.~Buican, A.~Dymarsky, and R.~Radhakrishnan, {\slshape {Quantum codes, CFTs, and defects},} \href{http://dx.doi.org/10.1007/JHEP03(2023)017}{{\em JHEP} {\bfseries 03} (2023) 017}, \href{http://arxiv.org/abs/2112.12162}{{ arXiv:2112.12162~[hep-th]}}.

\bibitem{Buican:2023ehi}
M.~Buican and R.~Radhakrishnan, {\slshape {Qudit stabilizer codes, CFTs, and topological surfaces},} \href{http://dx.doi.org/10.1103/PhysRevD.110.085021}{{\em Phys. Rev. D} {\bfseries 110} (2024) 085021}, \href{http://arxiv.org/abs/2311.13680}{{ arXiv:2311.13680~[hep-th]}}.

\end{thebibliography}\endgroup


\end{document}